\documentclass[12,aps]{revtex4}

\usepackage{pstricks}%
\usepackage{graphicx}
\usepackage{dcolumn}
\usepackage{bm}
\usepackage{dsfont}
\usepackage{amsmath}
\usepackage{pifont}
\usepackage{psfrag}

\renewcommand\Re{\text{Re\,}}
\renewcommand\Im{\text{Im\,}}
\newcommand\Tr{\text{Tr\,}}

\newcommand{\BE}{\begin{equation}}
\newcommand{\EE}{\end{equation}}

\newcommand{\tab}[1]{Tab.~\ref{#1}}
\newcommand{\fig}[1]{Fig.~\ref{#1}}
\newcommand{\eq}[1]{Eq.~(\ref{#1})}
\newcommand{\Sec}[1]{Sec.~\ref{#1}}

\newcommand{\App}[1]{Appendix~\ref{#1}}

\newcommand{\I}{\ensuremath{{\mkern1mu\mathrm{i}\mkern1mu}}}
\newcommand{\E}{\ensuremath{{\mkern1mu\mathrm{e}\mkern1mu}}}

\newcommand{\sep}{\ensuremath{{\mkern1mu\mathrm{sep}\mkern1mu}}}
\newcommand{\ps}{\ensuremath{{\mkern1mu\mathrm{ps}\mkern1mu}}}
\newcommand{\bs}{\ensuremath{{\mkern1mu\mathrm{bs}\mkern1mu}}}

\newcommand{\ket}[1]{\ensuremath{|#1\rangle}}
\newcommand{\bra}[1]{\ensuremath{\langle#1|}}
\newcommand{\braket}[2]{\ensuremath{\langle #1|#2\rangle}}

\begin{document}

\title{Revealing quantum properties with simple measurements}
\author{S.~W\"olk}
\affiliation{Department Physik, Naturwissenschaftlich-Technische Fakult\"at, Universit\"at Siegen - 57068 Siegen, Germany}

\begin{abstract}
Since the beginning of quantum mechanics, many puzzling phenomena which distinguish the quantum from the classical world, have appeared such as complementarity, entanglement or contextuality. All of these phenomena are based on the existence of non-commuting observables in quantum mechanics. Furthermore, theses effects generate advantages which allow quantum technologies to surpass classical technologies. 

In this lecture note, we investigate two prominent examples of these phenomenons: complementarity and entanglement. We discuss some of their basic properties and introduce general methods for their experimental investigation. In this way, we find many connections between the investigation of complementarity and entanglement. One of these connections is given by the Cauchy-Schwarz inequality which helps to formulate quantitative measurement procedures to observe complementarity as well as entanglement. 
\end{abstract}

\maketitle

\section{Introduction}
The quantum world inhabits many features such as complementarity \cite{Bohr1928}, entanglement \cite{Horodecki2009} or contextuality \cite{Kochen1967,Bell1966}  which distinguish it from the well known classical world. These very puzzling features have been the starting point of many discussions from the very first beginning of quantum mechanics. Furthermore, they are also the key ingredients of the advantages offered by quantum technology nowadays. Therefore, getting an understanding of these features, as deep as possible, is important to understand the foundations of quantum mechanics as well as  developing new applications of quantum technologies.

The most famous of these phenomenons is doubtlessly entanglement, which was first discussed by Einstein, Podolski and Rosen \cite{Einstein1935} and later quantified by Bell \cite{Bell1964}. After hearing the first time about entanglement, some people have the impression that entanglement is equal to anticorrelation between the measurement results of two parties. The reason for this misunderstanding is the famous example of the Bell state $\ket{\psi_-}=(\ket{01}-\ket{10})/\sqrt{2}$ where a spin-measurement in $z$-direction  on particle A reveals exactly the opposite measurement result than a measurement on particle B. However, this behavior has nothing to do with entanglement. First of all, the Bell state $\ket{\phi^+}=(\ket{00}+\ket{11})/\sqrt{2}$ is also entangled but leads to perfect correlation of spin measurements in $z$-direction on particle $A$ and $B$. Second, perfect anticorrelation also exist for classical measurements. Assume for example a box with a big sock and a small  sock. The socks are randomly distributed between Alice and Bob. A measurement of the size (big $\equiv 1$, small $\equiv -1$)  also leads to perfect anticorrelation. Therefore, anticorrelation alone is not an indicator for entanglement or quantumness in general.

In general, a measurement in a single measurement-basis alone can never demonstrate  quantum properties such as entanglement, discord, coherence or duality, because these measurement can always be simulated by a classical system. Quantum behavior can only be proven by measuring at least two observables which do not commute. In the classical world different observables always commute and are jointly measurable. Consecutively, different correlation between measurements $A_j$ and $B_j$ can influence each other as we illustrate on our example with the socks. Here, we use the measurement of the color (blue $\equiv 1$, red striped $\equiv -1$) as a second measurement-basis. We asume e.g. that with a certain probability $p_1$ the big sock is blue and the small one is red striped and with probability $p_2$ it is the other way around such that  $\langle A_1 B_1\rangle = \langle A_2 B_2\rangle = \langle A_2 B_1 \rangle  =-1/\sqrt{2}$. By fixing these three correlations, the fourth correlation is  bounded by
\BE
-\frac{1}{\sqrt{2}}\leq \langle A_1B_2 \rangle \leq 0 
\EE
 as summarized in \tab{tab:bell}. 

In quantum physics, not all observables are jointly measurable, e.g.  the observables described by the Pauli-spin matrices $\sigma_x^A \otimes \sigma_x^B$ and $\sigma_z^A\otimes \sigma_z^B$ are jointly measurable but $\sigma_z^A\otimes \sigma_x^A$ is not jointly measurable with the previous two. Therefore, the correlations  $\langle \sigma_x^A\sigma_x^B\rangle$ and $\langle \sigma_z^A\sigma_z^B\rangle$ can exist at the same time. However, they destroy correlation $\langle \sigma_z^A\sigma_x^B\rangle$.

As an example, we choose the observables $A_1=\sigma_x$, $A_2=\sigma_z$ and $B_1=(\sigma_x+\sigma_z)/\sqrt{2}$, $B_2=(\sigma_z-\sigma_x)/\sqrt{2}$ and the state $\ket{\psi_-}$. In this case, the resulting correlations  $\langle A_1 B_1\rangle = \langle A_2 B_2\rangle = \langle A_2 B_1 \rangle  =-1/\sqrt{2}$  are the same as in the classical example. However, because these three observables are not jointly measurable, they do not bound the correlation $\langle A_1 B_2\rangle$ as in the classical way and we get for our example   $\langle A_1 B_2\rangle = +1/\sqrt{2}$ which is classically forbidden (see \tab{tab:bell}).

A more general formulation of the connection between different correlations is given by the Bell inequalities such as the CHSH-inequality \cite{Clauser1969}
\BE
|\langle A_1B_1+A_2B_2+A_2B_1-A_1B_2\rangle|\underset{\sep}{\leq} 2 \label{eq:Bell}
\EE
which limits the correlation between classical observables.  However, it can be violated by using non-commuting observables and entangled states.

\begin{table}
\caption{Comparison of classical and quantum correlations. The observable are defined in the following way:(i) $A_1=B_1=$ size and $A_2=B_2=$ color for the classical case,  (ii) $A_1=\sigma_x$, $A_2=\sigma_z$ and $B_1=(\sigma_x+\sigma_z)/\sqrt{2}$, $B_2=(\sigma_z-\sigma_x)/\sqrt{2}$ for the quantum state. The correlations $\langle A_1B_1\rangle,\langle A_2B_2\rangle$  and $\langle  A_2B_1\rangle$ bound the allowed correlations $ \langle A_1B_2\rangle$ in the classical case but not in the quantum case.   }
\label{tab:bell}
\begin{tabular}{|c|c|c|c|c|}
\hline
state & $\langle A_1B_1\rangle$ & $\langle A_2B_2\rangle$ & $\langle  A_2B_1\rangle$ &$ \langle A_1B_2\rangle$\\ \hline
$p_1$ \begin{minipage}{0.07\textwidth}\includegraphics[width=1\textwidth]{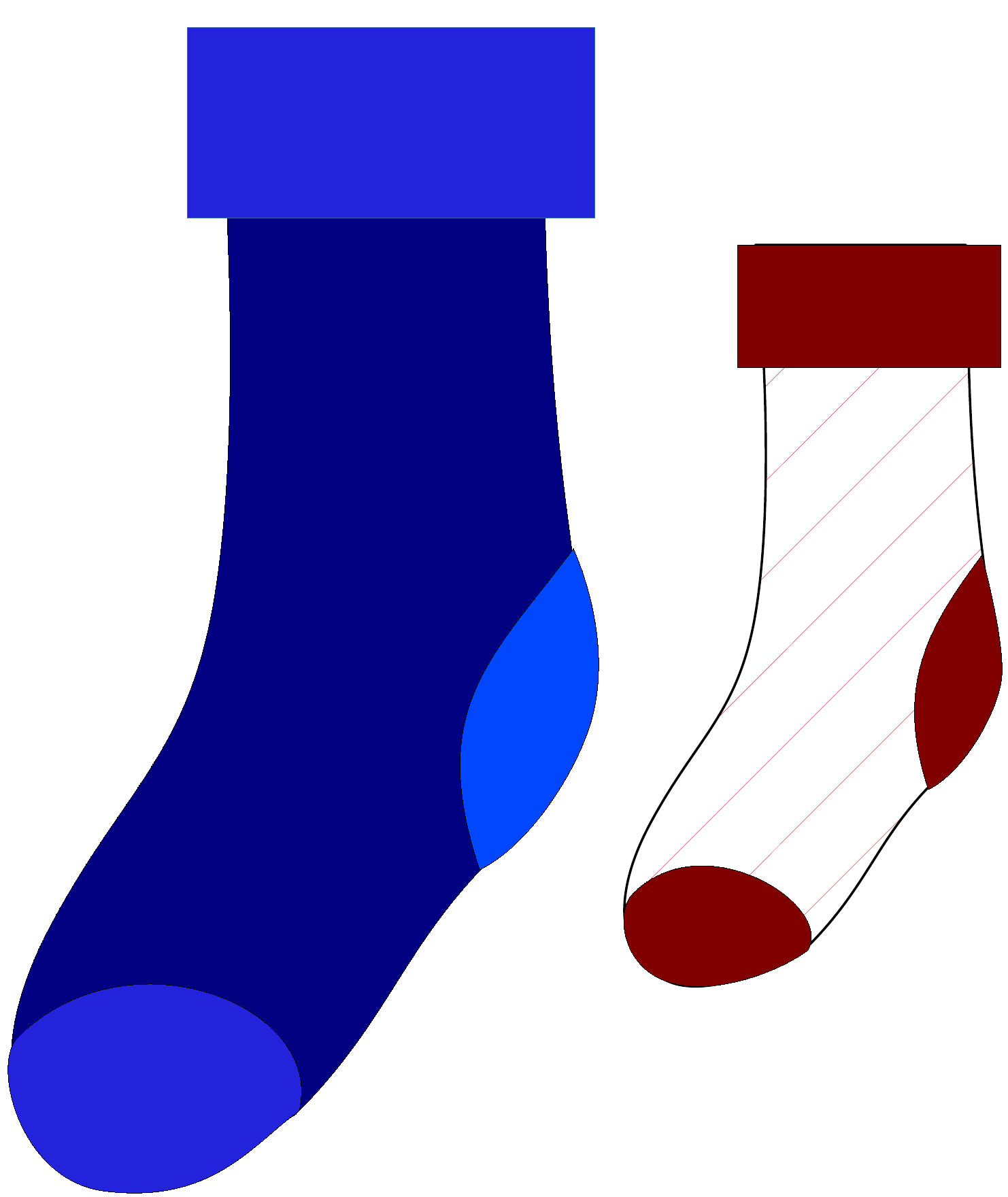}\end{minipage}+$p_2$ \begin{minipage}{0.07\textwidth}\includegraphics[width=1\textwidth]{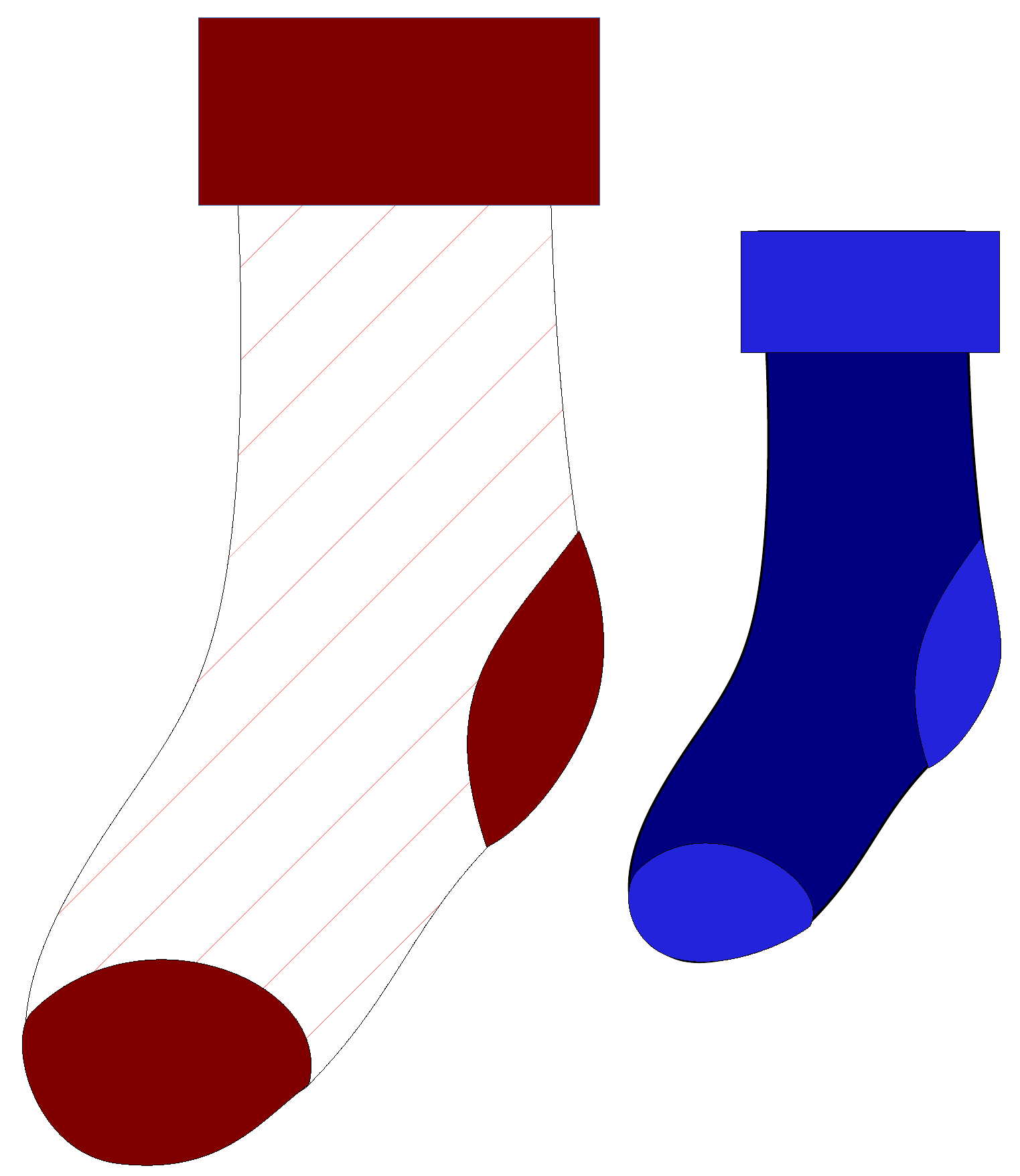}\end{minipage}& $-1/\sqrt{2}$& $-1/\sqrt{2}$&$-1/\sqrt{2}$&$ -1/\sqrt{2} \leq \dots \leq 0$\\[0.5em]
$\ket{\psi_-}=(\ket{01}-\ket{10})/\sqrt{2}$ & $-1/\sqrt{2}$ & $-1/\sqrt{2}$ & $-1/\sqrt{2}$ & $1/\sqrt{2}$\\ \hline
\end{tabular}
\end{table}

The phenomenon of non-commuting observables lies at the very heart of quantum mechanics and does not only induce the phenomenon of entanglement but also other quantum phenomena such  duality and uncertainty \cite{Zurek1979,Greenberger1988,Englert1996, Schleich2016},  contextuality \cite{Kochen1967} or discord \cite{Ollivier2002}. 

The aim of this lecture note is to give students with basic knowledge of quantum mechanics a better understanding of these quantum phenomena. Therefore, we will concentrate on a few selected topics, wave-particle duality as an example of complementarity (in \Sec{sec:wp-duality}) and entanglement (in \Sec{sec:entanglement}), rather than trying to give a complete picture.

In detail, we introduce in \Sec{sec:wp_in} the basic observables for wave-like and particle-like behavior and discuss the quantitative formulation of the wave-particle duality. Then, we give an overview of more advanced topics of wave-particle duality such as the quantum eraser (\Sec{sec:sim_meas}) and higher order wave-particle duality (\Sec{sec:higher_order}). We finish this chapter by pointing out the connection between measurements of wave-particle duality and entanglement in \Sec{sec:duality_ent}.

We start our investigation of entanglement (\Sec{sec:entanglement}) by summarizing the basic definitions and entanglement criteria for bipartite entanglement in \Sec{sec:bipartite}. Then, we discuss in detail tripartite entanglement in \Sec{sec:tripartite} and illustrate why tripartite entanglement is more than just the combination of bipartite entanglement. We finish this section by a brief overview on multipartite entanglement in \Sec{sec:multipartite} and a discussion on the spatial distribution of entanglement in \Sec{sec:ent_width}.
    
During the whole article, we will pay special attention  on how these quantum features can be quantified and observed. Of special interested is the Cauchy-Schwarz inequality which helps to quantify duality as well as entanglement with a few simple measurements.  


\section{Wave-particle duality\label{sec:wp-duality}}

Wave-particle duality is closely related to the question of what is light, which is one of the key questions that inspired the development of quantum mechanics.
Already the old Greeks discussed the nature of light, if light is a stream of particles (Democritus) or a ray (Plato, Aristotle). In the 17th century, the description of light as a wave (Huygens) made the construction of high quality lenses possible.  In the 18th and 19th century, the wave-theory of light was very successful to describe and explain refraction, diffraction and interference (Fresnel, Young).

However, the wave theory of light completely failed to explain effects such as the Compton effect, black body radiation or the photo effect investigated in the late 19th century  and the beginning of the 20th century. The contradiction of the wave- and the particle-nature of light might be best express by Einstein:  

``It seems as though we must use sometimes the one theory and sometimes the other, while at times we may use either. We are faced with a new kind of difficulty. We have two contradictory pictures of reality; separately neither of them fully explains the phenomena of light, but together they do''\footnote{Einstein, source: en.wikipedia.org/wiki/Wave-particle{\_}duality
}

The solution of this dilemma is given by the concept of complementarity best described by:

``If information on one complementary variable is in principle
available even if we choose not to ‘know’ it \dots we will
lose the possibility of knowing the precise value of the
other complementary variable.''\cite{Scully2007}.

Two complementary variable are the which-way information $D$ and the fringe visibility $V$ in a double-slit experiment or an interferometer. The complementarity principle can than be formulated in a quantitative way by
\BE
D^2+V^2 \leq 1 \label{eq:DVinequality}
\EE 
with $0\leq D,V\leq 1$. This inequality was first derived by D.M. Greenberger and A. Yasin for a specific way of defining the which-way information in an neutron interferometer \cite{Greenberger1988} before it was proven in a general way by B.-G. Englert \cite{Englert1996}.  

In the following, we will first discuss in \Sec{sec:wp_in} the two observables of which-way information and fringe visibility and how to measure and interpret them  before we derive \eq{eq:DVinequality}. Then, we discuss the generalization of this inequality to pairs of entangled particles in \Sec{sec:sim_meas} leading to the phenomenon of the so called quantum eraser\cite{Scully1981,Scully2000}.  Finally, we discuss wave-particle duality in the multi-photon case in \Sec{sec:higher_order}. Although, the wave-particle inequality \eq{eq:DVinequality} is valid in the mulit-photon case, it is very often not very informative. Therefore, we introduce a generalizations of \eq{eq:DVinequality} to the multi-photon case in \Sec{sec:higher_order}.


\subsection{Wave-particle duality: an inequality\label{sec:wp_in}}

\begin{figure}[t]
\begin{center}
\includegraphics[width=0.75\textwidth]{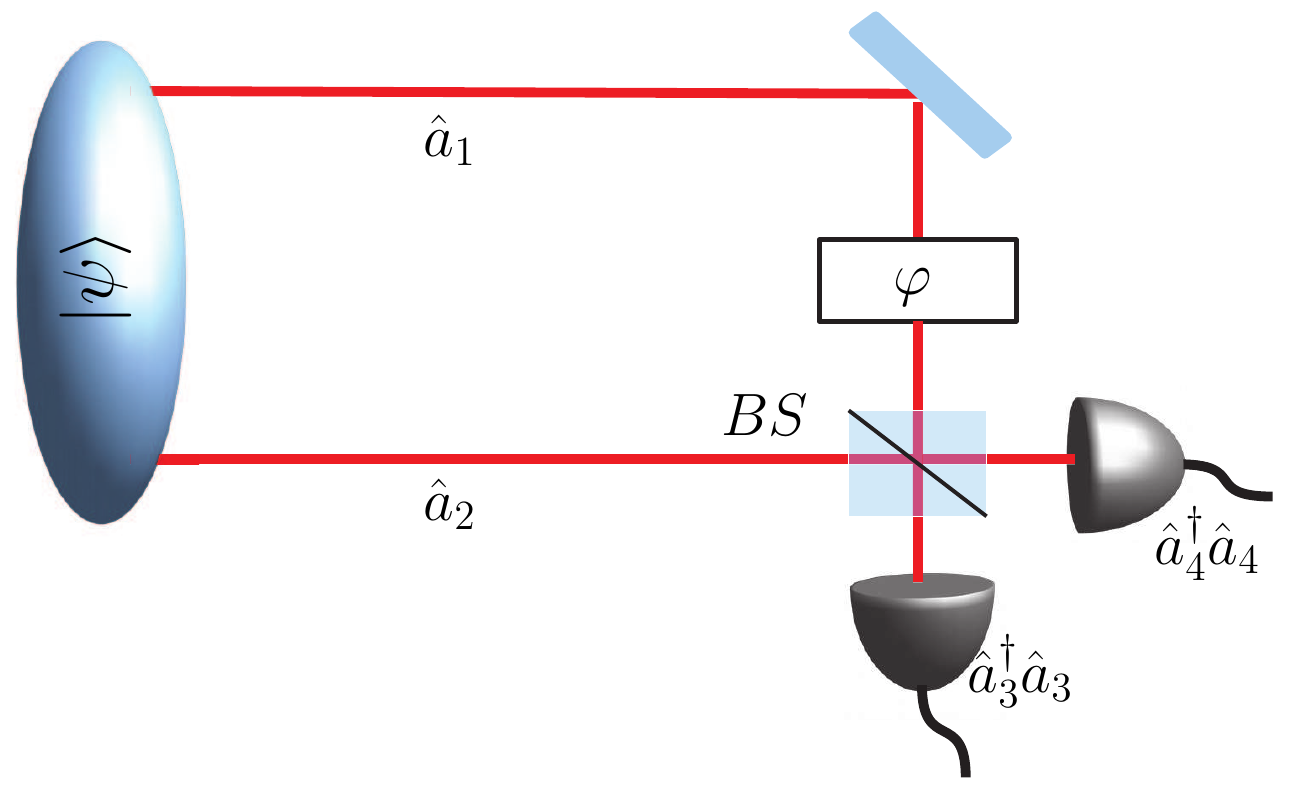}  
\end{center}   
\caption{General two mode state $\ket{\psi}$ of an interferometer. Different measurements can be performed by changing the phase $\varphi$ between the two paths and by inserting or excluding a beam splitter. }
\label{fig:interferometer}
\end{figure}

In the following, we will investigate the properties of a two mode quantum state $\ket{\psi}$ with a measurement setup given in \fig{fig:interferometer}. $\ket{\psi}$ can be described by
\BE
\ket{\psi}=f(a_1^\dagger, a_2^\dagger)\ket{0}_1 \ket{0}_2
\EE
with the creation operator $a_j^\dagger$ acting on mode $j$. The main properties of the creation and annihilation operator are given by
\begin{eqnarray}
[a_j,a_k^\dagger]&=&\delta_{j,k}\\
a^\dagger \ket{n-1}&=&\sqrt{n}\ket{n}\\
a\ket{n}&=&\sqrt{n}\ket{n-1}
\end{eqnarray}
where $\ket{n}$ describes a quantum state consisting of $n$ photons. In the following, we assume that we are able to perform two different kind of measurements (i) a simple photon number measurement in both modes described by their mean values
\BE
n_j=\langle \hat n_j\rangle = \langle a^\dagger_j a_j \rangle
\EE
or (ii) an interferometric measurement as shown in \fig{fig:interferometer}. Here, first the phase between both modes is changed via a unitary time evolution described by
\BE
U(\varphi)=\exp[-\I\varphi \;a^\dagger a].
\EE
Then, both modes described by $a_1$ and $a_2$ interact with each other via a $50:50$ beam splitter leading to the output modes
\BE
a_3=\frac{1}{\sqrt{2}}(a_1+a_2), \quad a_4=\frac{1}{\sqrt{2}}(a_1-a_2). 
\EE
Finally, a measurement of the photon number in each output mode is performed.  The description of the beam splitter is not unique and depends on the experimental realization of the beam splitter. However, all results obtained in this manuscript are valid independent of the actually used beam splitter transformation.

These two measurements can be used to measure the particle-like behavior, in form of the distinguishability,  and the wave-like behavior, in form of the visibility. These two observables can then by used to quantify the wave-particle duality as we will show in this section.


\subsubsection{Distinguishability}
If the photon behaves like a particle, it can only be in one of the two modes at the same time. If we always prepare the same quantum state,  we find $|\langle a_1^\dagger a_1 - a_2^\dagger a_2\rangle|=1$. As a consequence, we quantify the particle-like behavior via the distinguishability $D=|\langle \hat D\rangle|$ with 
\BE
\hat D \equiv \frac{a_1^\dagger a_1 - a_2^\dagger a_2}{\langle a_1^\dagger a_1\rangle + \langle a_2^\dagger a_2\rangle }
\EE 
which corresponds to a measurement setup as shown in \fig{fig:interferometer} without the beam splitter.

In other description of wave-particle duality, D describes the which-way information we gained by a measurement of a state $\ket{\tilde{\psi}}$ before the state reaches the beam splitter \cite{Englert1996}. However, in this case our state $\ket{\psi}$ just describes the state $\ket{\tilde{\psi}}$ after it was disturbed by the first measurement. In this case, the first which-way measurement is just part of the preparation of the state and the measurement described in our scenario is just a second measurement which will reveal exactly the same value if nothing happened in between.

In both cases $D$ described the probability of guessing the path of the photon correctly either after the fist measurement \cite{Englert1996} or in the next experiment. Here, $D=1$ corresponds to perfect knowledge and the photon will be always found in the same arm of an interferometer, whereas $D=0$ corresponds to no knowledge at all and the photon will be detected in both arms with equal probability.


\subsubsection{Visibility\label{sec:visibility}}

\begin{figure}[t]
\begin{center}
\includegraphics[width=0.4\textwidth]{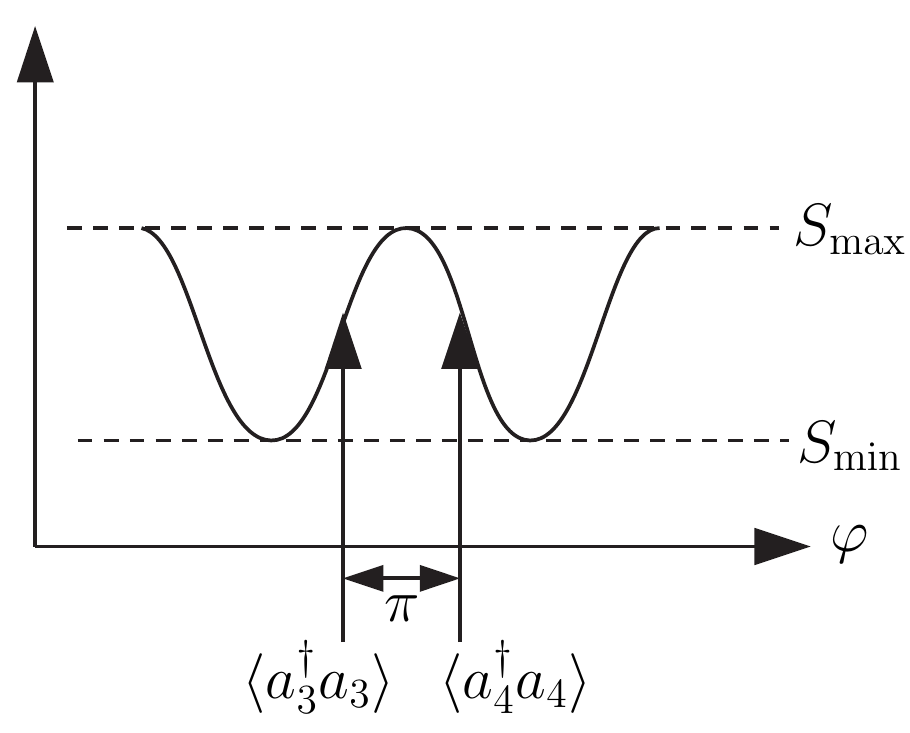}
\end{center}
\caption{Definition of fringe visibility in terms of the observables $\langle a_3^\dagger a_3\rangle_\varphi$ and $\langle a_4^\dagger a_4\rangle_\varphi$. The two observables depend on the variable $\varphi$ and $\langle a_4^\dagger a_4\rangle_\varphi$ correspond to $\langle a_3^\dagger a_3\rangle_{\varphi+\pi/2}$ shifted by $\pi$.   }
\label{fig:def_V}
\end{figure}

One important property of waves is that they can interfere with each others whereas particles cannot. The quality of an interference signale $S(\varphi)$ as depicted in \fig{fig:def_V} can be measured by the visibility $V$ which is given be the normalized difference between the maximum and the minimum of the signal
\BE
V=\frac{S_\text{max}-S_\text{min}}{S_\text{max}+S_\text{min}}.
\EE
A value of $V=1$ corresponds to perfect fringe visibility and perfect wave-like behavior whereas $V=0$ indicates that no fringes at all are visible.
If the phase shift and the beam splitter are present in  \fig{fig:interferometer}, then the two observables $\langle a_3^\dagger a_3\rangle$ and $\langle a_4^\dagger a_4\rangle$ correspond to exactly two measurement points of the interference signal separated by $\delta \varphi=\pi$. Therefore, the  fringe visibility is given by $V=\underset{\varphi}{\text{max}} \langle \hat V\rangle$ with 
\BE
\hat V \equiv \frac{a_3^\dagger a_3 - a_4^\dagger a_4}{\langle a_3^\dagger a_3\rangle + \langle a_4^\dagger a_4\rangle }.
\EE
However, we want to describe the fringe visibility as a property of the initial state. As a consequence, we need to rewrite $\hat V$ in terms of the input modes of the beam splitter. With the  help of the transformations
\begin{eqnarray}
a_3&=&\frac{1}{\sqrt{2}}(a_1\E^{-\I\varphi}+a_2)\\
a_4&=&\frac{1}{\sqrt{2}}(a_1\E^{-\I\varphi}-a_2)
\end{eqnarray}
we arrive finally at
\BE
\hat V = \frac{ a_1^\dagger a_2 \E^{\I\varphi} + a_1 a_2^\dagger  \E^{-\I\varphi} }{\langle a_1^\dagger a_1 \rangle + \langle a_2^\dagger a_2 \rangle}.
\EE
As a consequence, the fringe visibility is given by
\BE
V=\underset{\varphi}{\text{max}} \langle \hat V\rangle = \frac{2|\langle a_1^\dagger a_2\rangle|}{\langle a_1^\dagger a_1 \rangle + \langle a_2^\dagger a_2 \rangle}. 
\EE


\subsubsection{The wave-particle inequality}
After defining  a measure of particle-like and wave-like behavior, the main question is if we can observe both at the same time. The complementarity principle forbids $D=1$ and $V=1$ at the same time. However, what happens if we have some but not perfect path information, that is $0<D<1$. How much fringe visibility is possible in this case? This question can be answered by a simple calculation. The sum of the squares of both observables is given by
\BE
D^2+V^2 = 1-4\frac{\langle a_1^\dagger a_1\rangle \langle a_2^\dagger a_2\rangle- |\langle a_1^\dagger a_2\rangle|}{\langle a_1^\dagger a_1 \rangle + \langle a_2^\dagger a_2 \rangle}.
\EE
The Cauchy-Schwarz inequality given by
\BE
| \braket{\phi}{\tilde{\phi}}|\leq \braket{\phi}{\phi} \braket{\tilde{\phi}}{\tilde{\phi}}
\EE
leads to
\BE
\langle a_1^\dagger a_1\rangle \langle a_2^\dagger a_2\rangle- |\langle a_1^\dagger a_2\rangle| \geq 0
\EE
if we identify $\bra{\phi}=\bra{\psi}a_1^\dagger$ and $\ket{\tilde{\phi}}=a_2\ket{\psi}$. This inequality is also true for mixed states, which can be proven either with the convexity of $D$ and $V$ or again with the Cauchy-Schwarz inequality. As a consequence, we finally arrive at 
\BE
D^2+V^2 \leq 1 \label{eq:DV}
\EE
quantifying the wave-particle duality.
 Similar to the Heisenberg uncertainty relation, this inequality needs to be understood as a preparation inequality. That means, independent of how we measure $D$ and $V$, there exists no state $\rho$ which can be prepared in such a way that $D^2+V^2>1$. This is reflected by the fact that in the derivation of the inequality, we assumed that either the distinguishability or the visibility will be measured in each single run of the experiment, but never both at the same time. However, there exist also setups, with the goal of simultaneous measurements of $D$ and $V$ with the help of e.g. entangled photons. For a short overview, see \Sec{sec:sim_meas}.

In \tab{tab:state_examples} we have determined the distinguishability and and the visibility for different states. $\ket{\psi_1}$ and $\ket{\psi_2}$ are examples of two pure states with perfect particle-like behavior. The state  $\rho=p\ket{\psi_1}\bra{\psi_1}+(1-p)\ket{\psi_2}\bra{\psi_2}$ is an example of mixed state. The distinguishability $D^2=(2p-1)^2$ reduces with the mixedness $\gamma=\Tr \rho ^2=D^2/2+1/2$. However, no fringe visibility is gained by mixing.

$\ket{\psi_3}$ is an example of a state with perfect fringe visibility. In general, every pure one photon state can be written in the form of $\ket{\psi_4}$. By tuning the variable $\alpha$ any amount of $D$ or $V$ can be establish. However, for a given $\alpha$, the inequality \eq{eq:DV} is always saturated. A further investigation of $\ket{\psi_4}$ reveals, that a pure one-photon state leading to fringe visibility is always entangled and  in this case $V$ is a measure of entanglement (for further details about entanglement see \Sec{sec:entanglement}). However, the relation between fringe visibility and entanglement is only true for single-photon states. If several photons might be present, than also product states such as e.g. $\ket{\psi_7}$ are able to produce interferences fringes. In general, \eq{eq:DV} is also valid for multi-photon states as exemplified by the lower 4 examples in \tab{tab:state_examples}. However, in many cases the inequality does not contain much information because both observables are equal to zero and it is not possible to distinguish entangled states from separable states. For example the product state $\ket{\psi_7}$ and the mixed entangled state $\rho=\frac{1}{4} \ket{\psi_1}\bra{\psi_1}+\frac{1}{4} \ket{\psi_2}\bra{\psi_2}+\frac{1}{2} \ket{\psi_3}\bra{\psi_3}$ inhabit the same values for $D$ and $V$. However, also for multi-photon states, meaningful duality inequalities can be establish as we will show in \Sec{sec:higher_order}.

\begin{table}
  \caption{Distinguishability and visibility for different two-mode states}
  \label{tab:state_examples}
	\begin{tabular}{|c|c|c|}
\hline
State& $D_1$ & $V_1$   \\ \hline
$\ket{\psi_1}=\ket{01}$&1&0\\
$\ket{\psi_2}=\ket{10}$&1&0\\
$p\ket{\psi_1}\bra{\psi_1}+(1-p)\ket{\psi_2}\bra{\psi_2}$& $(2p-1)^2$&0\\
$\ket{\psi_3}=\left(\ket{1,0}+\ket{0,1}\right)/\sqrt{2}$&0&1 \\
$\ket{\psi_4}=\left(\cos \alpha \ket{1,0}+\sin \alpha\ket{0,1}\right)/\sqrt{2}$&$\cos^2 (2\alpha)$&$\sin^2(2\alpha)$ \\
$\ket{\psi_5}=\ket{1,1}$& 0 & 0\\
$\ket{\psi_6}=\ket{2,0}+\ket{0,2}$&0&0\\
$\ket{\psi_7}=(\ket{0}+\ket{1})(\ket{0}+\ket{1})$&0&1/2\\
$\frac{1}{4} \ket{\psi_1}\bra{\psi_1}+\frac{1}{4} \ket{\psi_2}\bra{\psi_2}+\frac{1}{2} \ket{\psi_3}\bra{\psi_3}$&0&1/2\\ \hline
\end{tabular}
\end{table}

\subsection{Simultaneous measurements \label{sec:sim_meas}}

\begin{figure}[t]
\begin{center}
\includegraphics[width=0.75\textwidth]{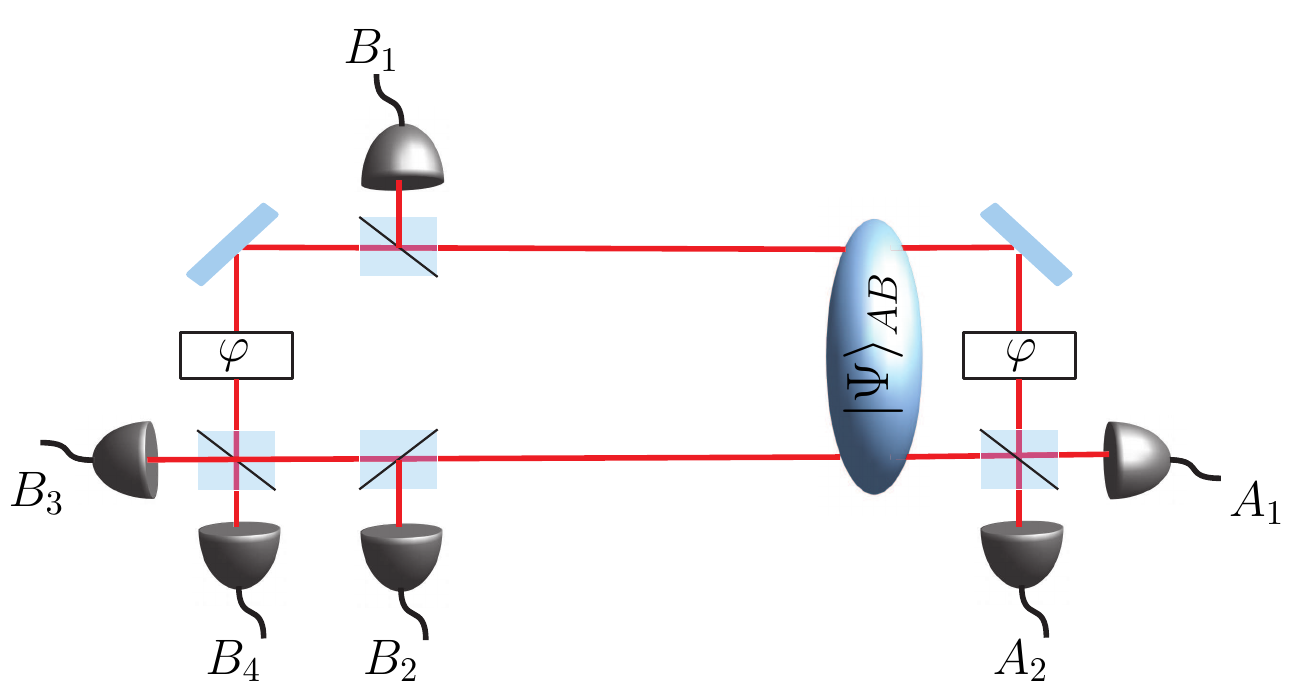}
\caption{Quantum eraser: Two entangled photons are both either prepared in the upper arm or the lower arm of the interferometer. Photon $A$ travels to the right  and is detected at detector $A_1$ or $A_2$. Photon $B$ travels to the left and is detected at detector $B_1$,$B_2$,$B_3$ or $B_4$. Coincidence measurement of photon A with detecting photon $B$ at detector $3$ or $4$ lead to interference whereas click at $B_1$ or $B_2$ ``eraser'' the interference.}
\label{fig:eraser}
\end{center}
\end{figure}

In the previous section, we discussed the wave-particle duality in a simple scheme, where we either measured the distinguishability $D$ or the visibility $V$ or had at least a defined temporal order of the measurements. Similar to the EPR setup \cite{Einstein1935} where position and momentum of a particle are measured ``simultaneously'', we can also measure the distinguishability and the visibility simultaneously by using entanglement and a setup as shown in \fig{fig:eraser}.  To perform this task, we entangle our photon $A$ with another photon $B$ with respect to their path information. In this way, we are able to measure the interference of a photon $A$ and later decide if we want to measure also its which-way information with the help of photon $B$. If we decide to measure the which-way information, the inference pattern must be erased due to the wave-particle duality, whereas the interference survives if we decide to not to measure the which-way information. This phenomenon is known as delayed choice quantum eraser \cite{Scully1981,Scully2000}. On the first sight, this phenomenon occurs to be quite puzzling because it seems that the measurement at photon $B$ changes the past of photon $A$. However, a simple analysis of this setup reveals that nothing is really erased and that \eq{eq:DV} is also valid in this case.

To understand this phenomenon, we assume a simple setup as displayed in \fig{fig:eraser}. The photons $A$ and $B$ are prepared in a superposition $\ket{\Psi}_{AB}=(\ket{00}+\ket{11})/\sqrt{2}$ where either both photons appear in the upper or lower arm of the interferometer. The interferometer is built in such a way, that photon $A$ reaches the detector $A_1$ or $A_2$ before photon $B$ reaches any beamsplitter or detector. Furthermore, the phase $\varphi$ is chosen in such a way that detector $A_1$ ($B_3$) clicks if photon $A$ ($B$)  is in state $(\ket{+}=\ket{0}+\ket{1})/\sqrt{2}$ and detector $A_2$ ($B_4$) clicks for the state $\ket{-}=(\ket{0}-\ket{1})/\sqrt{2}$. By inserting the additional beamsplitters and detectors $B_1$ and $B_2$ the photon itself randomly chose if the which-way information (detector $B_1$ or $B_2$ clicks) or the interference (detector $B_3$ or $B_4$ clicks) is measured.

Let us first assume, that Alice (measuring photon $A$) and Bob (measuring photon $B$) do not communicate. What does Alice observe? In this case, the reduce state at Alice side is given by
\BE
\rho_A=\Tr_B(\ket{\Psi}_{AB}\bra{\Psi})=\frac{1}{2}(\ket{0}\bra{0}+\ket{1}\bra{1})
\EE
which is a completely mixed state. From table \tab{tab:state_examples} we know that the fringe visibility for this state is given by $V_A=0$. By rewriting the reduced state $\rho_A=(\ket{+}\bra{+}+\ket{-}\bra{-})/2$ we can also interpret this experiment in such a way, that Alice gets randomly the state $\ket{+}$ with fringe-visibility $V_+=1$ and the state $\ket{-}$ with $V_-=-1$. As long as Alice is not able to distinguish whether she gets the state $\ket{+}$ or $\ket{-}$ via post selection, she is not able to see fringe visibility.   The distinguishability $D_A$ would be also equal to zero if Alice would have decided to measure it. As a consequence, Alice can neither see fringes nor does she has any which-path information without communicating with Bob. 

If Alice and Bob communicate with each other, we have to rewrite the state $\ket{\Psi}_{AB}$ in the corresponding measurement basis depending on the measurement outcomes. We rewrite the state 
\BE
\ket{\Psi}_{AB}= \frac{1}{\sqrt{2}}(\ket{00}+\ket{11})=\frac{1}{2}[(\ket{+}+\ket{-})\ket{0}_B+(\ket{+}-\ket{-})\ket{1}_B]
\EE
for a measurement of distinguishability at Bobs side (detector $1$ or $2$ clicks) and a fringe visibility measurement at Alice side. As a consequence, a click at detector $B_1\hat{=}\ket{0}_B$ corresponds to an equal probability that either $A_1\hat{=}\ket{+}$ or $A_2\hat{=}\ket{-}$ clicks. The same  happens for a click at detector $B_2$. As a consequence, a click at $B_1$ or $B_2$ does not help Alice to postselect between $\ket{+}$ and $\ket{-}$.

However, if $B_3$ or $B_4$ clicks, we describe the state by
\BE
\ket{\Psi}_{AB}= \frac{1}{\sqrt{2}}(\ket{00}+\ket{11})=\frac{1}{\sqrt{2}}(\ket{++}+\ket{--}).
\EE
 As a consequence, if Alice selects only the measurement outcomes which correspond to a click at $B_3\hat{=}\ket{+}$ ($B_4\hat{=}\ket{-}$) at Bob sides, she gets a perfect fringe-visibility of $V_{A|B_3} =1$ ($V_{A|B_4} =-1$).

As a consequence, the wave-particle duality in the form of \eq{eq:DV} stays also valid in this scenario. Furthermore, it is not the which-way measurement at Bobs side which erases the fringe-visibility at Alice side. However, it is the interference measurement at Bobs side which helps Alice to post select her measurement results to make the fringes appear. 

In general, no violation of the wave-particle duality is possible if the experimental data is analyzed correctly. However, sometimes the post selection as described above is not that apparent in a real experiment \cite{Menzel2012}. In these cases a violation of \eq{eq:DV} can occur due to unfair sampling \cite{Menzel2013,Boyd2016}.


\subsection{Higher-order wave-particle duality\label{sec:higher_order} }
The wave-particle duality as formulated in \eq{eq:DV} is valid for all states independent of the involved number of photons. However, the last four examples of \tab{tab:state_examples} also reveal, that \eq{eq:DV} might be not very informative if higher photon numbers are involved. In these cases, the generalization of \eq{eq:DV} to higher orders photon numbers as demonstrated in this section come into play. 

\subsubsection{Higher-order distinguisability}
The idea of a higher order distinguishability of order $k$ is to be only sensitive to states with at least $k$ photons. Therefore, we define the $k$th-order distinguishability
\BE
\hat{D}_k\equiv \frac{(a_1^\dagger)^k a_1^k - (a_2^\dagger)^k a_2^k}{\langle (a_1^\dagger)^k a_1^k \rangle + \langle (a_2^\dagger)^k a_2^k \rangle}
\EE
where we have used the $k$th-order autocorrelation function. In contrast to the $k$th-moment of the number operator given by $(a^\dagger a)^k$, the measurement of $(a^\dagger)^ka^k$ requires a different measurement setup \cite{Allevi2012}. Therefore, it really takes  additional information into account, whereas the higher moments can be estimated by a detailed data analysis of the data already collected for the distinguishability  $D_1$.

Possible measurements can e.g. be performed by detectors which need $k$-times the energy of a single photon to be activated \cite{Hemmer2006} or by photon counting measurements with single-photon resolution.

\begin{figure}[t]
   \begin{center}
     \includegraphics[width=0.45\textwidth]{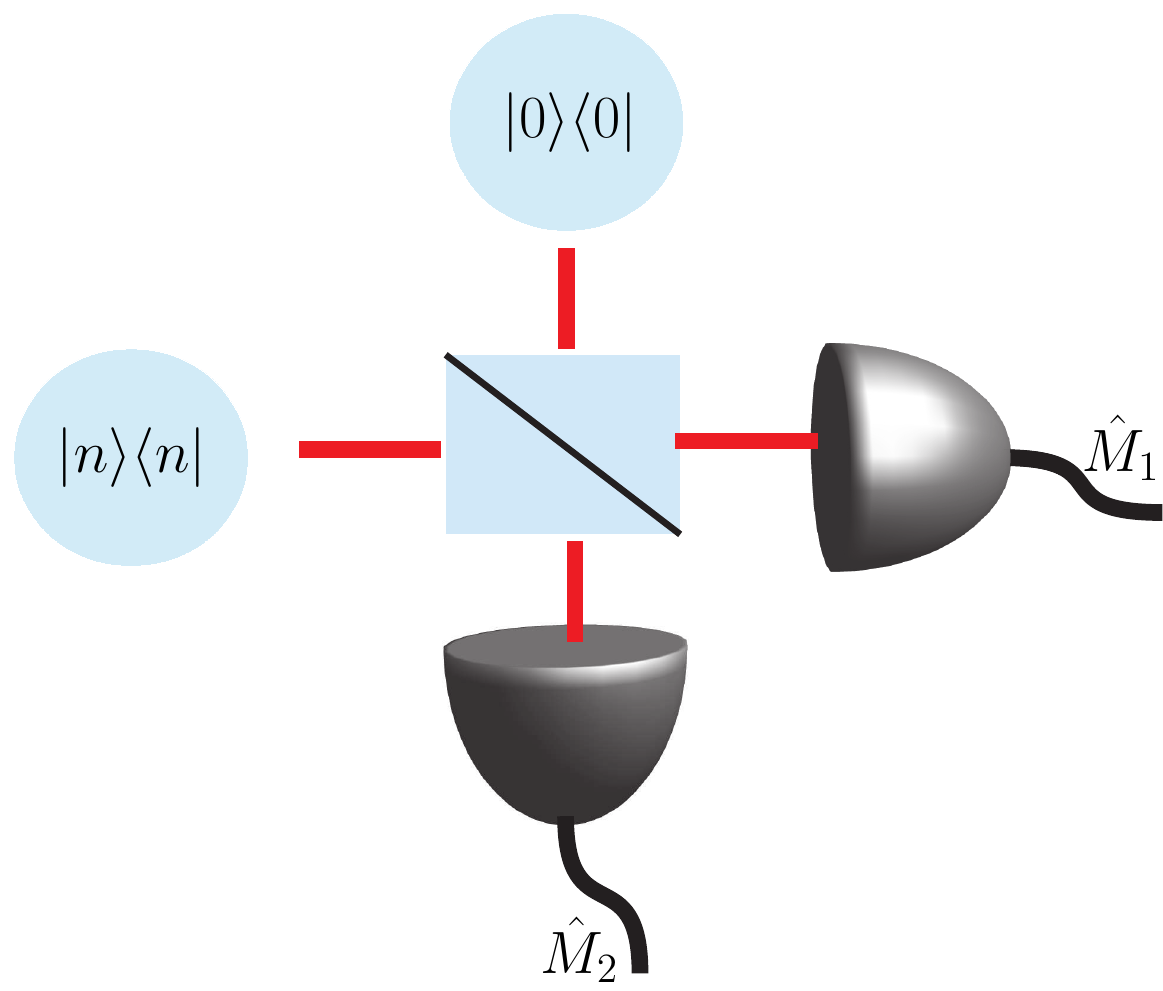}
		\end{center}
		\caption{The $k$th-order autocorrelation function can be interpreted as the probability to get exactly $k$ out of $n$ photons detector $M_1$. }
		\label{fig:autocorrelation}
\end{figure}

The action of the $k$th-order autocorrelation function on a Fock state $\ket{n}$ is given by
\BE
\frac{1}{k!} (\hat a^\dagger)^k \hat a^k \ket{n}={{n}\choose{k}}\ket{n}.
\EE 
As a consequence, the $k$th-order autocorrelation function is non-zero for states with $n>k$ photons. Contributions with different photon numbers $n$  are weighted with the binomial coefficient ${{n}\choose{k}}$. This fact can be best understood if we consider a measurement setup as given in \fig{fig:autocorrelation}. Here, a Fock state of exactly $n$ photons and the vacuum state interact on a 50:50 beam splitter. As a consequence, the probability that $k$ photons  leave the beam splitter at exit one is given by ${{n}\choose{k}}/\sqrt{2}$ similar to classical probability theory.  

\begin{figure}[t]
\begin{center}
\includegraphics[width=0.6\textwidth]{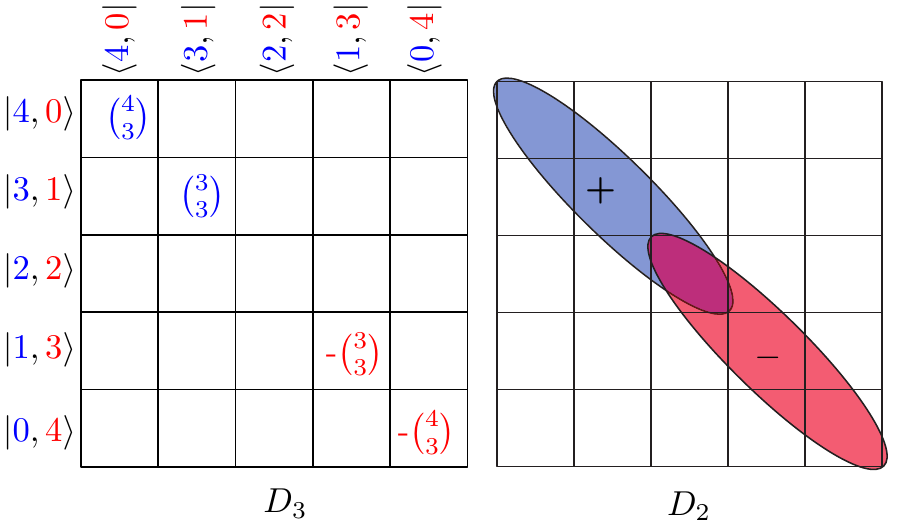}  
\end{center}   
\caption{The distinguishability $D_k$ is given by the sum of all diagonal terms with at least $k$ excitations in a single mode weighted with different weight factors depending on $k$ and the number of excitations.}
\label{fig:D_k}
\end{figure}

The $k$th-order distinguishability takes only the diagonal elements of the the state $\rho$ in the computational basis into account. We exemplify the different contribution for a state of constant total photon number $\langle a^\dagger_1 a_1+a^\dagger_2 a_2\rangle =4$ for $D_3$ and $D_2$ in \fig{fig:D_k}. 

\subsubsection{Higher-order visibility}
\begin{figure}[t]
\begin{center}
\includegraphics[width=0.35\textwidth]{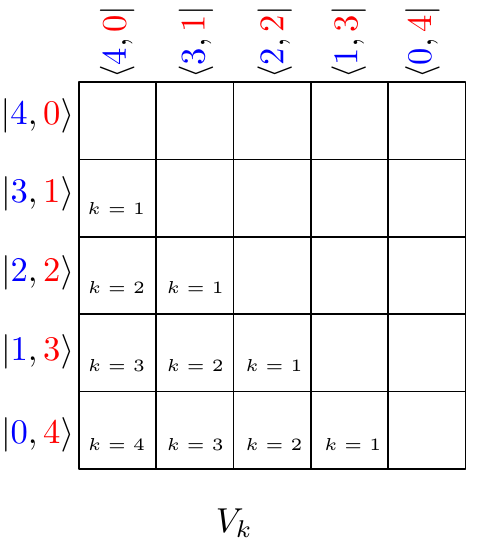}  
\end{center}   
\caption{The visibility $V_k$ is determined by the the off-diagonal elements of the state. Each order $k$ is given by another secondary diagonal. }
\label{fig:V_k}
\end{figure}

In a similar way, the fringes visibility $V$ can be generalized to
\BE
\hat V \equiv  \frac{(a_1^\dagger)^k a_2^k \E^{\I\varphi} + a_1^k  (a_2^\dagger)^k \E^{-\I\varphi}}{\langle (a_1^\dagger)^k a_1^k \rangle + \langle (a_2^\dagger)^k a_2^k \rangle}
\EE
where we have used the $k$th-order coherence
\BE
\frac{1}{k!}\bra{{n+k},{j}}{{(\hat a_1^\dagger)^k}} {{\hat a_2^k}}\ket{{{n}},{ j+k}}=\sqrt{{{n}\choose{k}}{{j}\choose{k}}}.
\EE
As a consequence, the $k$th-order visibility is sensitive to states where coherence exist between states where exactly $k$ photons change from one mode to the other. Therefore, $V_k$ is determined by the $k$th off-diagonal of $\rho$ in the computational basis as demonstrated in \fig{fig:V_k}.

Higher order visibilities $V_k$ cannot be measured by a single measurement setting in contrast to the first order visibility $V$. Nevertheless, the measurement setup depicted in \fig{fig:interferometer} together with detectors described by $(a^\dagger)^k a^k$ can determine $V_k$ if the measurements are performed for several different phases $\varphi_j$. We will demonstrate this measurement procedure for $k=2$. In this case, the sum of the expectation values of both detectors for an arbitrary but fixed phase $\varphi$ is given by
\begin{eqnarray}
R^+_{2,\varphi}&\equiv&\langle  (a_3^\dagger)^2 (a_3)^2+(a_4^\dagger)^2 (a_4)^2\rangle\\
&=& \frac{1}{2}\langle (a_1^\dagger)^2 a_1^2+(a_2^\dagger)^2 a_2^2+4a_1^\dagger a_1a_2^\dagger a_2 + \underbrace{(a_1^\dagger)^2 a_2^2 \E^{\I\varphi} + a_1^2  (a_2^\dagger)^2 \E^{-\I\varphi}}_{\sim\hat V_2} \rangle.
\end{eqnarray}
By using the correct phase relations, we are able to isolate the real and the imaginary part of $\langle (a_1^\dagger)^2 a_2^2 \rangle$ and arrive at
\BE
2|(a_1^\dagger)^k a_2^k |^2=  \underbrace{\left(R_{2,\varphi'}^+ - R_{2,\varphi'+\pi/2}^+\right)^2}_{\text{real part}} +  \underbrace{\left(R_{2,\varphi'-\pi/4}^+ - R_{2,\varphi'+\pi/4}^+\right)^2}_{\text{imaginary part}}.
\EE
In general, odd-orders of $V_k$ are determined by the difference of the detectors $M_3$ and $M_4$ whereas even-orders by their sum. By defining
\BE
R_{k,\varphi}^\pm\equiv \frac{2^{k-1}}{k}\langle  (a_3^\dagger)^k (a_3)^k\pm(a_4^\dagger)^k (a_4)^k\rangle,
\EE
we get for odd-orders of $k$
\BE
2|(a_1^\dagger)^k a_2^k |^2= \frac{2^{2k-2}}{k^2}\left[\left(\sum\limits_{m=0}^{k-1}R^-_{k,\varphi'+2m\pi/k}\right)^2 +  \left(\sum\limits_{m=0}^{k-1}R^-_{k,\varphi'-\pi/(2k)+2m\pi/k} \right)^2  \right]
\EE
and
\BE
2|(a_1^\dagger)^k a_2^k |^2= \frac{2^{2k-2}}{k^2}\left[\left(\sum\limits_{m=0}^{k-1}(-1)^m R^+_{k,\varphi'+m\pi/k}\right)^2 +  \left(\sum\limits_{m=0}^{k-1}(-1)^m R^+_{k,\varphi'-\pi/(2k)+m\pi/k} \right)^2 \right]
\EE
for even orders \cite{Woelk2013}.

														
\subsubsection{Higher-order wave particle duality}
The higher order distinguishability and visibility are again connected by the Cauchy Schwarz inequality
\BE
\left|\langle (a_1^\dagger)^k a_2^k\rangle\right|^2  \leq \langle (a_1^\dagger)^k a_1^k \rangle  \langle (a_2^\dagger)^k a_2^k \rangle
\EE
similiar to the first order wave particle duality. As a result, we get in the higher order case an exactly  similar duality relation given by
\BE
D_k^2+V_k^2 =1+4 \frac{\left|\langle(a_1^\dagger)^k a_2^k\rangle\right|^2  -\langle (a_1^\dagger)^k a_1^k \rangle  \langle (a_2^\dagger)^k a_2^k \rangle}{\left(\langle (a_1^\dagger)^k a_1^k \rangle + \langle (a_2^\dagger)^k a_2^k \rangle\right)^2} \leq 1
\EE
 as for the first order. Determining the higher order observables for the initial example states given in \tab{tab:state_examples} leads to the results shown  in \tab{tab:state_examples2}, which reveals the advantages of using a series of observables and duality relations instead of a single one. The here introduced higher order duality interprets a collection of $k$ photons as a single larger quasi particle. As a consequence, we get non-zero results for $D_2$ and $V_2$ for the states $\ket{\psi_6}=\ket{02}+\ket{20}$, $\ket{\psi_8}=\ket{2,0}$ and $\ket{\psi_9}$. The states $\ket{\psi_5}$ and $\ket{\psi_7}$ including also two photons states  lead to $ D_k=V_k=0$ on the other hand because the two photons do not bunch together.

\begin{table}
  \caption{Higher-order distinguishability and visibility for different two-mode states including the examples of \tab{tab:state_examples} which involve multi-photon states.}
  \label{tab:state_examples2}
	\begin{tabular}{|c|c|c|c|c|}
\hline
State& $D_1$ & $V_1$ &$D_2$ & $V_2$  \\ \hline
$\ket{\psi_5}=\ket{1,1}$& 0 & 0&0&0\\
$\ket{\psi_6}=\ket{2,0}+\ket{0,2}$&0&0&0&1\\
$\ket{\psi_7}=(\ket{0}+\ket{1})(\ket{0}+\ket{1})$&0&1/2&0&0\\
$\frac{1}{4} \ket{\psi_1}\bra{\psi_1}+\frac{1}{4} \ket{\psi_2}\bra{\psi_2}+\frac{1}{2} \ket{\psi_3}\bra{\psi_3}$&0&1/2&0&0\\
$ \ket{\psi_8}=\ket{20}$& 1&0&1&0 \\ 
$\ket{\psi_9}=\ket{4,2}+\ket{2,4}$&0&0&0&$(6/7)^2$\\
\hline

\end{tabular}
\end{table}

\subsection{Duality and entanglement\label{sec:duality_ent}}

\begin{figure}[t]
\begin{center}
\includegraphics[width=0.5\textwidth]{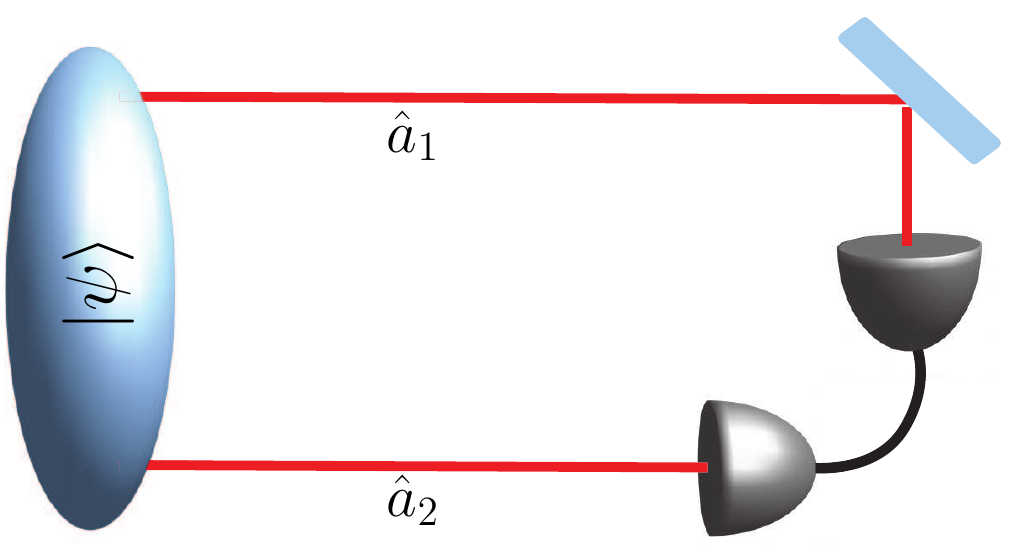}
\end{center}
\caption{Measurement setup of a coincidence measurement $C_k$ necessary to prove entanglement together with $V_k$.}
\label{fig:coincidence}
\end{figure}

The higher order visibility $V_k$ is similar to the first order visibility an indicator of entanglement if the total photon number of the state is fixed. However, for states with a varying total photon number additional measurements are necessary for proving entanglement. As we will demonstrate in \Sec{sec:entanglement} entanglement can be verified by comparing diagonal and off-diagonal elements with the help of the Cauchy-Schwarz inequality. Entanglement criteria can be developed by applying the Cauchy-Schwarz inequality (CSI) only to subsystems (for more details see \Sec{sec:entanglement}) whereas the duality inequality follows from applying the CSI to the total system. As a consequence, the higher-order visibility $\hat{V}_k$ needs to be compared to the higher order coincidence
\BE
\hat C_k\equiv \frac{(a_1^\dagger)^k a_1^k  (a_2^\dagger)^k a_2^k}{\left(\langle (a_1^\dagger)^k a_1^k \rangle + \langle (a_2^\dagger)^k a_2^k \rangle\right)^2},
\EE 
which can be measured by a setup similar to \fig{fig:coincidence}.
Separable states obey the inequality
\BE
V_k^2\underset{\sep}{\leq} 4C_k \label{eq:V_criterion}
\EE  
which follows directly from \eq{eq:Cauchy_sep}, which we will prove in the next section, by using $\hat A_1 = (a_1^\dagger)^k$, $\hat B_2 = a_2^k$ and $\hat A_2=\hat B_1= \mathds{1}$. Violation of this inequality demonstrates entanglement.

In general, also the distinguishability $D_k$ can be used to detect entanglement by using
the observable $W_k=\underset{\varphi}{\textrm{max}}|\langle \hat W_k\rangle|$ with
\BE
 \hat W_k \equiv \frac{a_1^k a_2^k \E^{\I\varphi} + (a_1^\dagger)^k (a_2^\dagger)^k \E^{-\I\varphi}}{\langle (a_1^\dagger)^k a_1^k \rangle + \langle (a_2^\dagger)^k a_2^k \rangle} 
.
\EE
Entanglement of a state is demonstrated if it violates the inequality
\BE
D_k^2 + W_k^2 \underset{\sep}{\leq}1,\label{eq:D_criterion}
\EE
which can be proven again by \eq{eq:Cauchy_sep}.

The entangled states of \tab{tab:state_examples} can all be detected by the criterion \eq{eq:V_criterion} as demonstrated in \tab{tab:state_examples3}. On the other side, \eq{eq:D_criterion} only detects states with coherence between $\ket{n,j}$ and $\ket{n+k,j+k}$ such as e.g. $\ket{\psi}=\sqrt{3}\ket{00}+\ket{11}$ with $D_1=0$ but $W_1=\sqrt{3}$.

\begin{table}
  \caption{Results of the entanglement test using \eq{eq:V_criterion} }
  \label{tab:state_examples3}
	\begin{tabular}{|c|c|c|c|c|c|}
\hline
State& $D_1$ & $V_1$ & $C_1$ & $W_1$ & entangled \\ \hline
$\ket{\psi_4}=\left(\cos \alpha \ket{1,0}+\sin \alpha\ket{0,1}\right)/\sqrt{2}$&$\cos^2 (2\alpha)$&$\sin^2(2\alpha)$ & 0& 0& \textrm{yes} \\
$\ket{\psi_5}=\ket{1,1}$& 0 & 0&1/4&0&\textrm{no}\\
$\ket{\psi_7}=(\ket{0}+\ket{1})(\ket{0}+\ket{1})$&0&1/2&1/16&0& \textrm{no} \\
$\rho=\frac{1}{4} \ket{\psi_1}\bra{\psi_1}+\frac{1}{4} \ket{\psi_2}\bra{\psi_2}+\frac{1}{2} \ket{\psi_3}\bra{\psi_3}$&0&1/2&0&0& \textrm{yes} \\ \hline
\end{tabular}
\end{table}

\section{Entanglement\label{sec:entanglement}}

Entanglement \cite{Horodecki2009} is the most puzzling and gainful concept of quantum mechanics and a direct consequence of the concept of superposition of quantum states. Originally this effect was formulated 1935 as a contradiction to reality obeying local realism by Einstein, Podolsky and Rosen \cite{Einstein1935}. It took 29 years until the difference between local realism and quantum mechanics could be quantified \cite{Bell1964} and experimentally tested by means of of the Bell inequality. Many experiments \cite{Clauser1972,Clauser1976,Fry1976,Zeilinger2015,NIST2015} have been performed since then, which strongly support the concept of quantum mechanics. Most importantly entanglement is used in quantum technologies such as quantum information processing \cite{Monroe2002} and quantum metrology \cite{Toth2014} to overcome limits of classical devices.

A deep understanding of entanglement in theory and experiment helps us to develop quantum technologies. Therefore, it is also important to develop entanglement criteria \cite{Guehne2009} which can be easily applied to experiments to certify and investigate the existing entanglement in a given setup and its behavior under the influence of noise. 

The most famous entanglement criterion is given by the Bell inequality in \eq{eq:Bell}.
However, not all entangled states violate the Bell inequalities. Moreover, many highly entangled states such as the Greenberger-Horne-Zeilinger (GHZ) state \cite{Greenberger1990} which violate  Bell inequalities are very vulnerable with respect to noise and therefore not the best choice for e.g. quantum metrology under realistic conditions \cite{Altenburg2016}. Thus, further criteria to characterize entanglement also for weakly entangled states are necessary. 

In the following, we will give a small review about entanglement with special attention given to the question of experimental entanglement characterization.  We first define bipartite entanglement and give several examples of how to detect it experimentally. Then, we will continue with tripartite and multipartite entanglement and demonstrate that multipartite entanglement is more than the summation of bipartite entanglement.


\subsection{Bipartite entanglement\label{sec:bipartite}}
 A pure quantum state is entangled, if it cannot be written as
\BE
\ket{\Psi}_{A,B}=\ket{\psi}_A \otimes \ket{\phi}_B;
\EE
a mixed quantum state is entangled, if it cannot be written as
\BE
\rho_{A,B}=\sum\limits_j p_j\ket{\psi_j}_A\bra{\psi_j} \otimes \ket{\phi_j}_B\bra{\phi_j}.\label{def:ent_mix}
\EE
A problem of this definition is that the decomposition of a mixed quantum state is not unique. Therefore \eq{def:ent_mix} indicates separability if such a decomposition is found. Though, a decomposition including entangled states does not prove that the mixed state itself is entangled. The definition of bipartite entanglement of pure states, on the other hand, can be used to directly demonstrate entanglement with the help of the Schmidt decomposition \cite{Ekert1995}.

\subsubsection{Schmidt decomposition}
The Schmidt decomposition \cite{Ekert1995} is a method to transform a bipartite pure state $\ket{\psi}_{A,B}$ into the following form
\BE
\ket{\Psi}_{A,B}=\sum\limits_{j=1}^r p_j \ket{a_j}\ket{b_j} \label{eq:Schmidt}
\EE
where $\lbrace \ket{a_j}\rbrace$ and $\lbrace \ket{b_j}\rbrace$ form an orthogonal basis of the two parties $A$ and $B$, respectively. The maximal summation index $r$ is called Schmidt rank and indicates entanglement if $r>1$. We will demonstrate the Schmidt decomposition on the example
\BE
\ket{\Psi}_{A,B}=\frac{1}{\sqrt{6}}\left(\ket{00}+\ket{01}+\sqrt{2}\ket{10}-\sqrt{2}\ket{11}\right).
\EE
This state is obviously  not in the form given by \eq{eq:Schmidt} because the basis states $\ket{0},\ket{1}$ appear twice for each party. To determine the correct basis states for the Schmidt decomposition, we have to determine the eigenstates of the reduced density matrix
\begin{eqnarray}
\rho_B&=&\Tr_A \left(\ket{\Psi}_{A,B}\bra{\Psi}\right)\\
&=& \frac{1}{3}\ket{+}\bra{+}+\frac{2}{3}\ket{-}\bra{-}
\end{eqnarray}
with $\ket{\pm}=(\ket{0}\pm\ket{1})/\sqrt{2}.$
The Schmidt coefficients are given by $p_j=\sqrt{\lambda_j}$ with the eigenvalues $\lambda_j$ of $\rho_B$. The Schmidt decomposition of the state $\ket{\Psi}_{A,B}$ is given by its decomposition into the eigenstates of $\rho_B$. Thus, the Schmidt decomposition of $\ket{\Psi}_{A,B}$ is given by
\BE
\ket{\Psi}_{A,B}=\sqrt{\frac{1}{3}}\ket{0}_A\ket{+}_B + \sqrt{\frac{2}{3}}\ket{0}_A\ket{-}_B,
\EE 
and  $\ket{\Psi}_{A,B}$ is obviously entangled since $r=2>1$. The Schmidt rank is independent of whether it is determined via the reduced density matrix of system $A$ or system $B$. In a simplified version, the Schmidt decomposition states that a pure bipartite state is entangled if its reduced density matrices are mixed. However, to apply the Schmidt decomposition the state has not only to be pure, but also we need full knowledge about the state. This is rather difficult and extensive measurements are required. Nevertheless, the Schmidt decomposition can be very helpful to identify  the optimal measurement directions for experimental entanglement verification \cite{Schwemmer2012} as we will demonstrate with the following example.

One way to detect entanglement of a two-qubit system is to measure various combinations of the Pauli matrices, that is
\BE
M_{j,k}=\sigma_j\otimes \sigma_k.
\EE
The sum of these observables is bounded from above by
\BE
\sum\limits_{j,k=1}^3 \langle M_{j,k} \rangle_\sep\leq 1 \label{eq:exp_schmidt}
\EE
for separable states, whereas this bound can be violated by entangled states (for proof see \App{app:exp_schmidt}). In the worst case, 9 observables are needed to be measured to detect entanglement via this criterion. Though, for pure entangled states, it is sufficient to measure two observables, if the right directions are chosen \cite{Schwemmer2012}. To find these directions, Alice and Bob need to determine their reduced density matrices by measuring their local bloch vectors defined by $\vec{v}=(\langle \sigma_x \rangle,  \langle \sigma_y \rangle, \langle \sigma_z\rangle)^T$. With the help of the eigenstates $\ket{v}$ and $\ket{v_\perp}$ of the operator $V=\vec{v}\vec{\sigma}$ they define the new measurement directions
\BE
\sigma_{z'}=\ket{v}\bra{v}-\ket{v_\perp}\bra{v_\perp},\; 
	\sigma_{y'}=\I \ket{v_\perp}\bra{v}-\I\ket{v}\bra{v_\perp}.
\EE
Every pure entangled state can then be detected by solely measuring $M_{z'z'}$ and $M_{y'y'}$ and using \eq{eq:exp_schmidt}. In contrast to the Schmidt decomposition, the states do not necessarily need  to be pure. As soon as we find $M_{z'z'}+M_{y'y'}>1$ we know that the state is entangled independent of whether the state is pure or not. The only difference is, that for strongly mixed states, maybe additional terms have to be measured to exceed the threshold. 

\subsubsection{The positive partial transpose criterion}

In a real quantum experiment, we can never guarantee pure quantum states. Furthermore, interesting phenomena of entanglement appear when we consider not only  pure but mixed states (see \Sec{sec:tripartite}). Therefore, it is very convenient to have entanglement criteria, which are also valid  for mixed states. The most famous criterion for entanglement of mixed states is the positive partial transpose (PPT) criterion \cite{Peres1996,Horodecki1996}. It is based on the fact, that the transposition is a positive but not completely positive map. This means that the transpose of $\rho_A$ is positive semidefinite ($\rho_A^T\geq 0$)  if $\rho_A$ itself is positive semidefinite. Still, if the transpose is only applied to a subsystem of a composed quantum state $\rho_{A,B}$, this is not necessarily the case. Assume for example the Bell state $\ket{\psi^+}_{A,B}=(\ket{00}+\ket{11})/\sqrt{2}$ and its corresponding density matrix 
\BE
\rho_{A,B}\equiv \frac{1}{2} \left(\ket{00}\bra{00}+\ket{11}\bra{00}+\ket{00}\bra{11}+\ket{11}\bra{11}\right).
\EE
Its partial transpose is given by 
\BE
\rho^{PT}_{A,B}= \frac{1}{2} \left(\ket{00}\bra{00}+\ket{01}\bra{10}+\ket{10}\bra{01}+\ket{11}\bra{11}\right)
\EE
with the resulting eigenvalues $\lambda=\pm 1/2$. 
In the matrix representation (here in the standard computational basis $\ket{00},\ket{01},\ket{10},\ket{11}$) the partial transposition is calculated by dividing the the matrix into 4 submatrices is seen below 
\BE
\left(\begin{array}{cc|cc}
1&0&0&1\\ 0&0&0&0 \\ \hline 0&0&0&0 \\ 1&0&0&1
\end{array}\right)
\; \underset{PPT}{\Rightarrow} \; 
\left(\begin{array}{cc|cc}
1&0&0&0\\ 0&0&1&0 \\ \hline 0&1&0&0 \\ 0&0&0&1
\end{array}\right).
\EE
The partial transpose is then calculated by either exchanging the two off-diagonal matrices as a total or by transposing each submatrix depending on whether the partial transpose should be calculated with respect to partition $A$ or $B$. This example shows that the partial transpose of an entangled state can lead to negative eigenvalues. We call states with negative partial transpose such as $\ket{\psi^+}$ NPT-states.

On the other hand, the partial transpose of a separable state
\BE
\rho^{PT}_\sep = \sum\limits_j p_j \ket{\psi_j}_A\bra{\psi_j}  \otimes \ket{\phi_j^\ast}_B\bra{\phi_j^\ast}
\EE
with $\rho_B^T=\rho_B^\ast$, stays a valid quantum state with positive eigenvalues. As a consequence, we find that a state is entangled if it possesses a negative partial transpose. Yet, this criterion is only necessary and sufficient for $2\times 2$ and $2\times 3$ systems \cite{Horodecki1996}. For all other systems, we can make no statement about the entanglement if a state is PPT. 

Nevertheless, if a state is PPT or not is a very important property. First of all, even if a state with PPT is entangled, it is only bound entangled \cite{Horodecki1998}. This means, even if Alice and Bob possess several copies of this state, they cannot create the singlet state $\ket{\psi^-}$ out of them by using only local operations and classical communication (LOCC). In addition, bipartite bound entangled states cannot be directly used for teleportation or densed coding. Though, they can support these tasks if used as additional resources \cite{Horodecki1999}. Furthermore, PPT can be much easier tested and can be used to detect multipartite entanglement with the help of semidefinite programming \cite{Hofmann2014}. As a consequence, the PPT criterion is a very important criterion for theoretical entanglement investigations. However, the experimental application of the PPT criterion suffers from several disadvantages. First of all, total knowledge of the state is necessary (gained e.g. by state tomography) which leads to great experimental effort. Furthermore, experimental state tomography can lead to negative eigenvalues of the state $\rho$ itself \cite{Moroder2013} if using linear inversion. The use of the maximum likelihood method to avoid this negativity can lead to an overestimation of the entanglement \cite{Schwemmer2015}. Hence, the most convenient way to experimentally verify entanglement is given by inequalities directly based on experimental observables as we will demonstrate in the next section.

\subsubsection{Detecting entanglement with the help of the Cauchy-Schwarz inequality}

The most convenient way to experimentally detect entanglement are inequalities of expectation values. There exist many different kinds of entanglement criteria based on inequalities, e.g. for discrete and continuous variables \cite{Shchukin2005,Hillery2006} in the  bipartite and multipartite case \cite{Hillery2010,Guehne2010,Huber2010}. Originally, they have been developed with different methods such as using uncertainty relations or using the properties of separable states.  Although all the above cited criteria did not seem to have much in common on the first sight, all of them can be proven with a single general method \cite{Woelk2014}. This general method is given by the Cauchy-Schwarz inequality as we will demonstrate in this section.

The expectation value of an operator can be upper bounded by the Cauchy-Schwarz inequality by interpreting it as the scalar product of two vectors. In general, an operator $O$ can be part of the bra- or the ket-vector or can be split into $O=O_1O_2$. Therefore, the most general bound is given by
\BE
|\underbrace{\bra{\psi} O_1}_{\vec{x}^\dagger}\underbrace{O_2\ket{\psi}}_{\vec{y}}|^2 \leq \underbrace{\bra{\psi}O_1O_1^\dagger \ket{\psi}}_{|\vec{x}|^2}\underbrace{\bra{\psi}O_2^\dagger O_2\ket{\psi}}_{|\vec{y}|^2} .
\EE  
Hence, we get the general bound 
\BE
|\langle A_1A_2B_1B_2\rangle|^2 \leq \langle A_1A_1^\dagger B_1B_1^\dagger \rangle \langle A_2^\dagger A_2 B_2^\dagger B_2\rangle \label{eq:CS_general}
\EE
for a bipartite state  where the operators $A$ and $B$ acting solely on subsystem $A$ or $B$, respectively.
The expectation value $\langle AB\rangle$ factorizes for product states (ps). As a result, we find the upper bound
\begin{eqnarray}
 | \langle   A_1   A_2  B_1   B_2\rangle|^2 & \underset{\ps}{=} &|\langle A_1A_2\rangle||\langle B_1B_2\rangle|^ \\
&\underset{\ps}{\leq}& \langle A_1A_1^\dagger\rangle \langle A_2A_2^\dagger\rangle \langle B_1B_1^\dagger\rangle \langle B_2B_2^\dagger\rangle
\end{eqnarray}
for product states. Expectation values  of different subsystems can be recombined again, which leads us to the general entanglement criterion 
\BE
 | \langle   A_1   A_2  B_1   B_2\rangle|^2\underset{\sep}{\leq}  \langle   A_1  A_1^\dagger   B_2^\dagger  B_2\rangle\langle   A_2^\dagger  A_2 B_1  B_1^\dagger\rangle. \label{eq:Cauchy_sep}
\EE 
This inequality is also valid for mixed separable states \cite{Woelk2014}. If \eq{eq:Cauchy_sep} provides a stricter bound than \eq{eq:CS_general}, then it can be used to detect entanglement. For example, the criterion \cite{Hillery2006}
\BE
 |\langle a^m (b^\dagger)^n\rangle|^2\underset{\sep}{\leq}\langle  (a^\dagger)^m a^m  (b^\dagger)^n b^b\rangle,
\EE
where $a$ and $b$ denote the annihilation operators of system $A$ and $B$, follows directly by choosing $A_2= a$, $B_1=b^\dagger$ and $A_1=B_2=\mathds{1}$. The entanglement criterion
\BE
|\rho_{01,10}|^2 \underset{\sep}{\leq} \rho_{00,00}\rho_{11,11},
\EE
with $\rho_{jk,nm}$ denoting the matrix entries of a two-qubit state, follows from $A_1=B_2=\ket{0}\bra{0}$ and $A_2=B_1^\dagger=\ket{0}\bra{1}$. 

The best lower bound for separable states is achieved by using operators of the form $A_1=\ket{a}\bra{\varphi}$ and $A_2=\ket{\varphi}\bra{\alpha}$ with $\ket{a},\ket{\alpha}$ and $\ket{\varphi}$ being arbitrary states of system $A$ and likewise choosing the operators $B_1$ and $B_2$. The criterion is independent of the state $\varphi$ and linear combinations of such operators lead to weaker bounds \cite{Woelk2014}. In the bipartite case, the entanglement criterion given in \eq{eq:Cauchy_sep} detects only states with NPT and it is necessary and sufficient for two-qubit states if we optimize over all measurement directions. In addition, it detects all NPT states of the form
\BE
\rho=p \ket{\psi}\bra{\psi}+ \frac{1-p}{d}\mathds{1}_d.
\EE
The choice of the best operator given by $A_1A_2=\ket{a}\bra{\alpha}$  with two arbitrary but different states $\ket{a},\ket{\alpha}$ leads to a measurement of mean values of non-hermitian operators. As a consequence, $|\langle A_1A_2B_1B_2\rangle|$ cannot be measured directly. Nevertheless, the expectation value of a non-hermitian operator $O$ can be estimated via the two hermitian operators
\begin{eqnarray}
2\Re[\langle O \rangle] &=& \langle O+O^\dagger\rangle,\\
2\Im[\langle O \rangle] &=& \langle \I O^\dagger-\I O\rangle.
\end{eqnarray}
In photon experiments, the real and imaginary part can be measured e.g. with the help of a beam splitter (compare \Sec{sec:visibility}). In experiments with trapped ions, the off-diagonal terms can be obtained by measurements in the basis
\BE
\ket{\leftarrow}=(\ket{0}+\E^{\I\phi}\ket{1})/\sqrt{2}, \quad \ket{\rightarrow}=(\ket{0}+\E^{\I(\phi+\pi)}\ket{1})\sqrt{2}.
\EE
The parity of the measurement given by the probability that the blochvectors of both ions point in the same direction
\BE
\Pi(\varphi)=P_{\leftarrow,\leftarrow}+P_{\rightarrow,\rightarrow}
\EE
then determines the off diagonal elements via $2|\rho_{00,11}|=\Pi_\textrm{max}-\Pi_\textrm{min}$ .


\subsection{Tripartite entanglement\label{sec:tripartite}}
One might think first, that multipartite entanglement can be characterized by investigating the entanglement of all possible bipartite splits. However, this is only true for pure states. Mixed entangled states cannot be fully characterized by this attempt.

\begin{figure}[t]
\begin{center}
\includegraphics[width=0.6\textwidth]{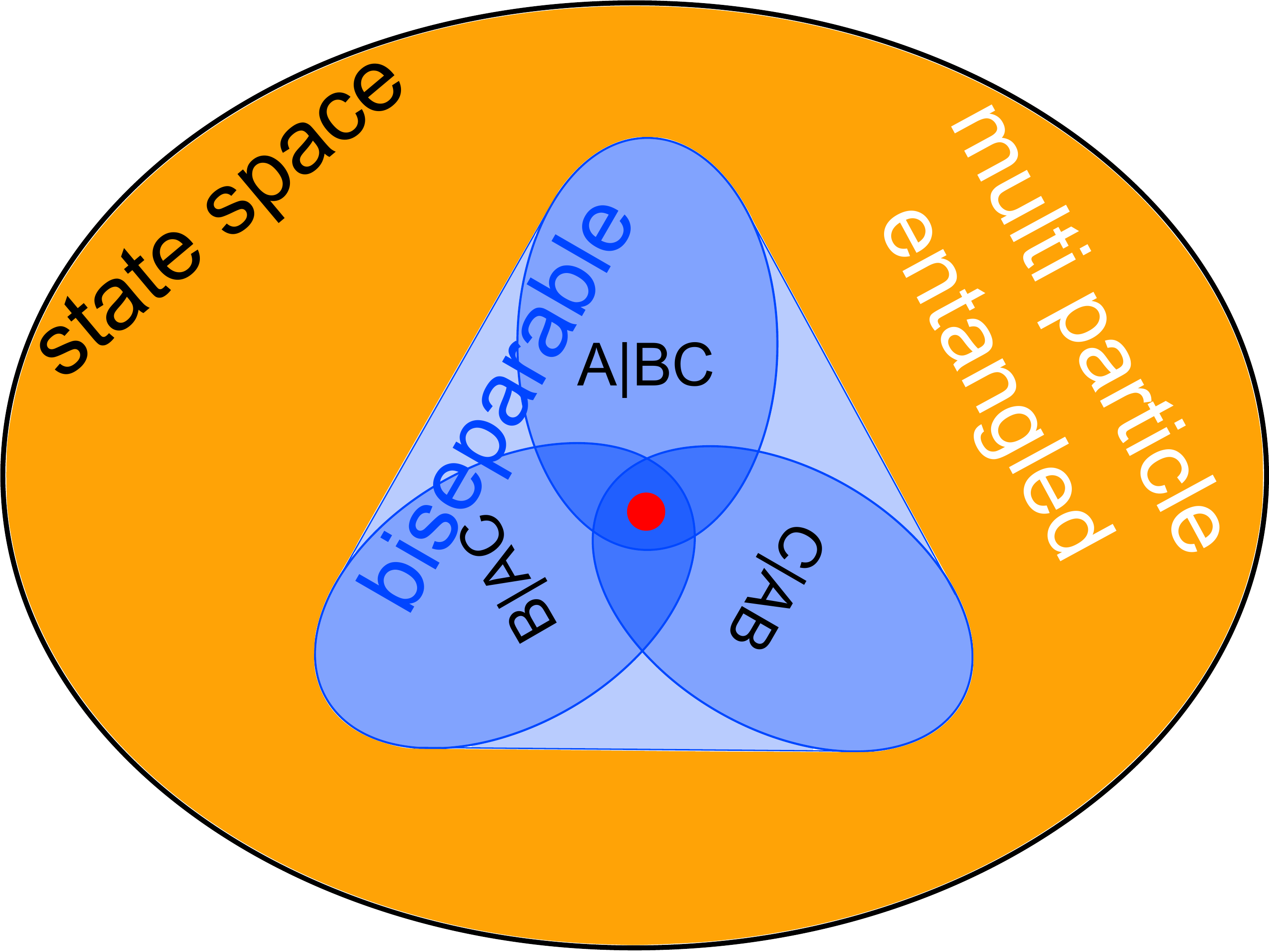}
\end{center}
\caption{Demonstration of the characterization of entanglement in a tripartite systems. The inner red dot represents the fully separable states which is a subset of the biseparable states (inner blue triangle). States which lie outside of the triangle are genuine multipartite entangled. There exist different kinds of bipartite states: (i) bipartite states with respect to every bipartition (overlapped of all three ellipses), (ii)  bipartite states with respect to a given bipartition (single ellipse), (iii) convex combination of bipartite states (convex triangular hull)\cite{Hofmann2014}.}
\label{fig:3_ent}
\end{figure}

There exist 4 different groups of tripartite states. If $A,B$ and $C$ denoting the three partition, then the different groups can be characterized in the following way \cite{Cirac2000}:
\begin{enumerate}
\item Fully separable states
\BE
\rho_\sep = \sum\limits_j p_j\; \rho_j^A \otimes
\rho_j^B \otimes \rho_j^C.
\EE
\item Biseparable states with respect to a given partition, e.g. $A|BC$
\BE
\rho_{A|BC}=\sum\limits_j p_j \;\rho_j^A \otimes
\rho_j^{BC}.
\EE
\item General bipartite states, which are convex combination of bipartite states with respect to different bipartitions
\BE
\rho_\bs= p_1 \;\rho_{A|BC}+p_2\; \rho_{AB|C}+p_3 \; \rho_{AC|B}\label{eq:bisep}.
\EE
\item Genuine multipartite entangled states given by all states which cannot be represented by \eq{eq:bisep}.
\end{enumerate}
So, on the one hand there exist states which are biseparable under every bipartition but not fully separable. On the other hand, there also exist states which are not separable under any given bipartition, but are still not genuine multipartite entangled. The different types of tripartite states form a nested structure consisting of convex groups as shown in \fig{fig:3_ent}.

Entanglement in the tripartite case can be also detected with the help of inequalities of expectation values. In Ref. \cite{Hillery2006}, the authors suggested that their entanglement criterion can be generalized to the multipartite case by
\BE
|\langle \prod\limits_{k=1}^{N}  O_k\rangle|^2 \underset{\sep}{ \leq}  \langle \prod\limits_{k=1}^j  O_k^\dagger   O_k  \prod\limits_{k=j+1}^{N}   O_k   O_k^\dagger\rangle, \label{eq:Zubairy_gen1}
\EE
with $ O_k$ being an operator acting on system $k$. Still, this generalization just divides all parties into the two subgroups $\lbrace 1,2,\dots, j\rbrace$ and $ \lbrace j+1,\dots,N\rbrace$. As a result, it does not investigate real multipartite entanglement but only bipartite entanglement with respect to a given bipartition.

A more advanced generalization is given by
\BE
 |\langle \prod\limits_{k=1}^n  O_k\rangle|\underset{\sep}{\leq}\prod\limits_{k=1}^n\langle (  O^\dagger_k O_k)^{n/2} \rangle^{1/n}, \label{eq:Hillery1}
 \EE
which is not equivalent to checking a given single bipartition. Nevertheless, it cannot reveal the complex structure of tripartite entanglement because it is still based on combinations of bipartite entanglement \cite{Woelk2014}. For example, \eq{eq:Hillery1} is not only valid for separable states, but also for states were for each pair of subsystems $k$ and $j$ we find a bipartition $M | \bar M$  such that $k \in M$ and $j \in \bar M$ and the state
$\rho$ is biseparable for this bipartition $M | \bar M$. In the tripartite case, this means that e.g. states which are biseparable under the partition $A|BC$ and $B|AC$ do not violate \eq{eq:Hillery1}.

A more complete characterization of tripartite entanglement can be achieved with the criteria developed in Ref. \cite{Guehne2010}. Here, the entanglement of three-qubit states are investigated by developing entanglement criteria based on the matrix entries $\rho_{j,k}$ given in the standard product basis $\lbrace \ket{000},\ket{001},\dots,\ket{111}\rbrace$. General biseparable states (bs) satisfy the inequalities
\begin{eqnarray}
|\rho_{1,8}|&\underset{\bs}{\leq}&\sqrt{\rho_{2,2}\rho_{7,7}}+\sqrt{\rho_{3,3}\rho_{6,6}}+\sqrt{\rho_{4,4}\rho_{5,5}},\\
|\rho_{2,3}|+|\rho_{2,5}|+|\rho_{3,5}|&\underset{\bs}{\leq}&\sqrt{\rho_{1,1}\rho_{4,4}}+\sqrt{\rho_{1,1}\rho_{6,6}}+\sqrt{\rho_{1,1}\rho_{7,7}}+\frac{1}{2}(\rho_{2,2}+\rho_{3,3}+\rho_{5,5}).
\end{eqnarray} 
Violation of at least one of the two inequalities indicates genuine multipartite entanglement. Fully separable states obey the inequalities
\begin{eqnarray}
|\rho_{1,8}|&\underset{\sep}{\leq}&(\rho_{2,2}\rho_{3,3}\rho_{4,4}\rho_{5,5}\rho_{6,6}\rho_{7,7})^{1/6},\label{eq:seev1}\\
|\rho_{1,8}|&\underset{\sep}{\leq}&(\rho_{1,1}\rho_{4,4}^3\rho_{5,5}\rho_{6,6}\rho_{7,7})^{1/6}.\label{eq:seev2}
\end{eqnarray}
These inequalities are also able to detect weak entangled states such as states which are bisarable under every bipartition but still not fully separable \cite{Guehne2010}. 

Criteria for genuine multipartite entanglement can be generated by convex combinations of biseparable criteria. There exists also a general method to derive entanglement criteria for weak entangled states such as multipartite states which are PPT under every bipartition. This general method is based on the H\"older inequality, which is a generalization of the Cauchy-Schwarz inequality. 

For product states, we factorize again the expectation value 
\BE
\langle A_1A_2B_1B_1C_1C_2\rangle \underset{\sep}{\leq} \sum\limits_j p_j |\langle A_1 A_2\rangle_j|\cdot |\langle B_1 B_2\rangle_j|\cdot |\langle C_1 C_2\rangle_j|,
\EE
where we have used that each separable state $\rho$ can be written as a convex combination $\rho=\sum_j p_j \ket{\psi_j}_\ps\bra{\psi_j}$ of product states. Consecutively, we 
 use the Cauchy-Schwarz inequality for each single subsystem $A$, $B$ or $C$ and recombine them again. At this point, it is also possible to increase the number of expectation values by using $\langle  OO^\dagger\rangle  = \sqrt[k]{\langle OO^\dagger\rangle^k}$. In this way, we find e.g. the inequality
\begin{eqnarray}
|\langle   A_1  A_2  B_1  B_2  C_1  C_2\rangle| 
\underset{\sep}{\leq} \sum\limits_j p_j&& \sqrt[4]{\langle   A_1  A_1^\dagger   B_1  B_1^\dagger   C_2^\dagger   C_2\rangle_j\langle   A_1  A_1^\dagger   B_2^\dagger  B_2   C_1   C_1^\dagger\rangle_j}  \nonumber\\
&&\times\sqrt[4]{ \langle   A_2^\dagger  A_2   B_1  B_1^\dagger   C_1   C_1^\dagger\rangle_j\langle   A_2^\dagger   A_2  B_2^\dagger  B_2   C_2^\dagger   C_2\rangle_j}.
\end{eqnarray}
 This inequality still depends on the representation $\rho=\sum_j p_j \ket{\psi_j}_\ps\bra{\psi_j}$ which is not unique. However, with the help of the generalized H\"older inequality
\BE
\sum\limits_j p_j x_j y_j\leq \Big(\sum\limits_j p_j x_j^{1/r}\Big)^r
\Big(\sum\limits_j p_j y_j^{1/s}\Big)^s\label{eq:hoelder}
\EE
and $\langle O \rangle =  \sum_j p_j \langle O\rangle$ we finally arrive at
\begin{eqnarray}
  |\langle   A_1  A_2  B_1  B_2  C_1  C_2\rangle|
  &\underset{\sep}{\leq}&\sqrt[4]{\langle   A_1  A_1^\dagger   B_1  B_1^\dagger   C_2^\dagger   C_2\rangle_\sep\langle   A_1  A_1^\dagger   B_2^\dagger  B_2   C_1   C_1^\dagger\rangle} \nonumber \\
	&& \times \sqrt[4]{\langle   A_2^\dagger  A_2   B_1  B_1^\dagger   C_1   C_1^\dagger\rangle_\sep\langle   A_2^\dagger   A_2  B_2^\dagger  B_2   C_2^\dagger   C_2\rangle} \label{eq:Cauchy4}.
\end{eqnarray}
This inequality cannot be derived by combining bipartite entanglement criteria and can therefore also detect PPT states such as

\BE
\varrho_\alpha=\frac{1}{8+8\alpha}\left(\begin{array}{cccccccc}
	                 4+\alpha &0&0&0&0&0&0&2 \\
	                 0&\alpha&0&0&0&0&2&0\\
	                 0&0&\alpha&0&0&-2&0&0\\
	                 0&0&0&\alpha&2&0&0&0\\
	                 0&0&0&2&\alpha&0&0&0\\
	                 0&0&-2&0&0&\alpha&0&0\\
	                 0&2&0&0&0&0&\alpha&0\\
	                 2 &0&0&0&0&0&0&4+\alpha \\
	                \end{array}
	                \right),
\EE
which is a PPT entangled state for $2\leq \alpha\leq 2\sqrt{2}$ \cite{Kay2011,Guehne2011, Woelk2014}. The entanglement of this state can be detected with the help of \eq{eq:Cauchy4} for $2\leq \alpha<2.4$.

The entanglement criteria \eq{eq:Hillery1}, \eq{eq:seev1} and \eq{eq:seev2} can be also derived with the  above described scheme. However, \eq{eq:Hillery1} and \eq{eq:seev1} cannot detect the entanglement of the state $\varrho_\alpha$  \cite{Woelk2014}.
\subsection{Multipartite entanglement \label{sec:multipartite}}

The exact characterization of tripartite entanglement includes already many different kinds of entanglement as we have seen in the previous section. Therefore, the investigation of mulit-partite entanglement concentrates on a few very characteristic properties instead of pursuing for a complete characterization which would be too complicated. In general, there exist different ways of categorizing multipartite entanglement: (i) $n$-separability and (ii) $k$-producibility / entanglement depth $k$ \cite{Molmer2001,Guehne2005}. 

The first categorization asks whether a pure state $\ket{\psi}$ can separated into the product of $n$ groups. As a result, we call a  pure state $\ket{\psi}$ $n$-separable if it can be written as
\BE
\ket{\psi}=\ket{\phi_1}\otimes\ket{\phi_2}\otimes \cdots \otimes \ket{\phi_n}.
\EE
A mixed state is $n$-separable, if it is a convex combination of $n$-separable pure states. In general, each of the states $\ket{\phi_j}$ can consist of an arbitrary number of parties not necessarily of equal dimension. Moreover, a $n$-separable state is also $n-1$ separable and so on. Thus, the declaration of a state as $n$-separable is in general only a lower bound on the separability if not declared different. To proof that a $N$-partite state is entangled, we need to proof that this state is not $N$-separable. For genuine multipartite entanglement, we need to show that is not biseparable.

\begin{figure}[t]
\begin{center}
\includegraphics[width=0.75\textwidth]{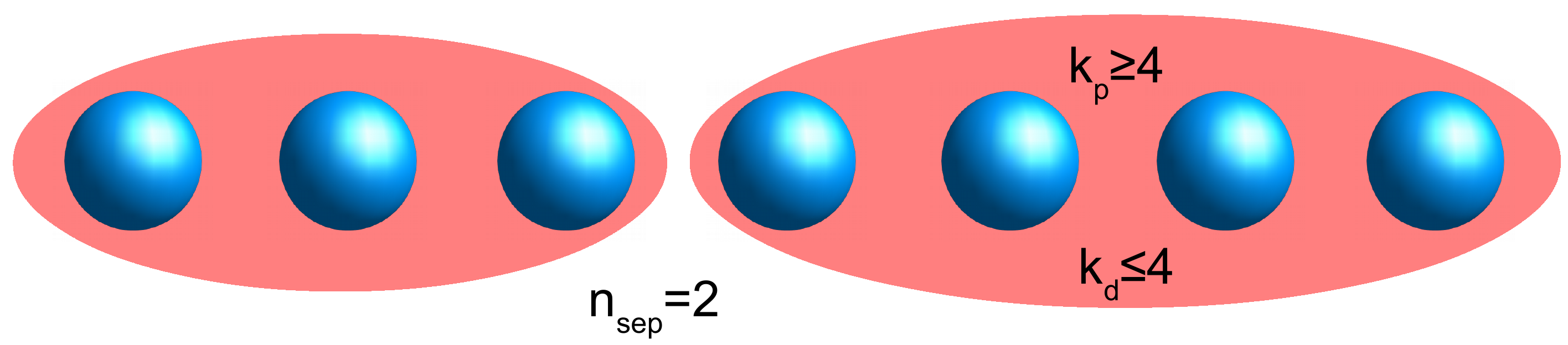}
\end{center}
\caption{Comparison of the entanglement categories: $n$-separability ($n_\sep$), $k$-producability ($k_p$) and entanglement depth ($k_d$).}
\label{fig:ent_depth}
\end{figure}

The second categorization does not ask about the number of groups but how many parties are in each group (see \fig{fig:ent_depth}). This question can be asked in two different ways: (i) the $k_p$-producability asks how many entangled particles are needed to create a state and (ii) the entanglement depth $k$ defines how much entanglement can be extracted from the state. On the one hand, a $k_p$-producible state is also $k_p+1$ producible  and so on. Hence, it is an upper bound of the actual entanglement. On the other hand, a state of entanglement depth $k_d$ inhabits also a depth of $k_d-1$ and so on. Therefore, the entanglement depth is a lower bound. We call a state $k$-partite entangled if $k_p=k_d=k$. 

Multipartite entanglement plays an important role in quantum metrology schemes \cite{Toth2014}. Here, the precision $(\triangle \theta)^2$ of determining an unknown parameter $\theta$ is lower bounded by 
\BE
(\triangle \theta)^2 \geq \frac{1}{sk^2+r}
\EE
for a $k$-partite entangled state of a total number of $N=sk+r$ parties \cite{Hyllus2012}.

Multipartite entanglement can be detected e.g. by measuring the global spin $\vec{J}=\sum_k \vec{\sigma}_k$ \cite{Luecke2014,Hosten2016}. For $k$-partite entangled states, the variance $(\triangle J_z)^2$ is limited by
\BE
(\triangle J_z)^2 \geq J_\text{max}\cdot  F_{k/2}\left( \frac{\sqrt{\langle J_x^2+J_y^2\rangle}}{J_\text{max}}\right),\label{eq:mulit_ent}
\EE
with the effective spin length $\sqrt{\langle J_x^2+J_y^2\rangle}$ and the maximal spin $J_\text{max}=N/2$ for spin $1/2$ particles. The function $F_{k/2}$ can be determined numerically by using that the state minimizing $(\triangle J_z)^2$  is symmetric and given by $\ket{\Psi}=\ket{\psi}^{\otimes N/k}$ \cite{Luecke2014}. In this way, entanglement of more than $680\pm 35$ $^{87}Rb$ atoms has been demonstrated experimentally \cite{Hosten2016}.

\subsection{The spatial distribution of entanglement\label{sec:ent_width}}

\begin{figure}[t]
\begin{center}
\includegraphics[width=0.5\textwidth]{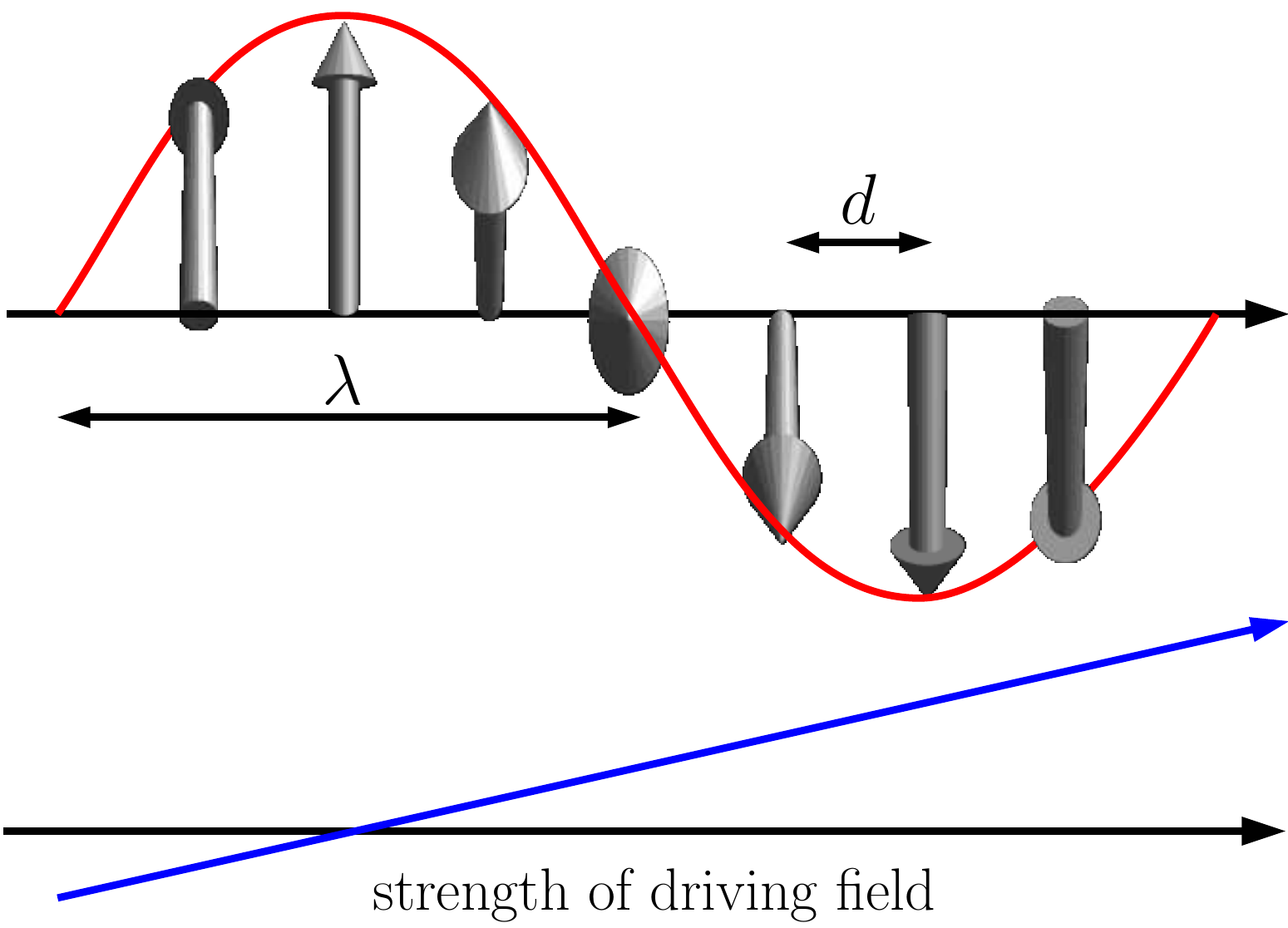}
\end{center}
\caption{The magentic gradient field (lower picture) give rise to a position dependent rotation of the spin (upper picture). The projection of the spin in one direction leads to a sinusoidal observable. }
\label{fig:spin_rotation}
\end{figure}

\begin{figure}[t]
   \begin{center}
   \includegraphics[width=1\textwidth]{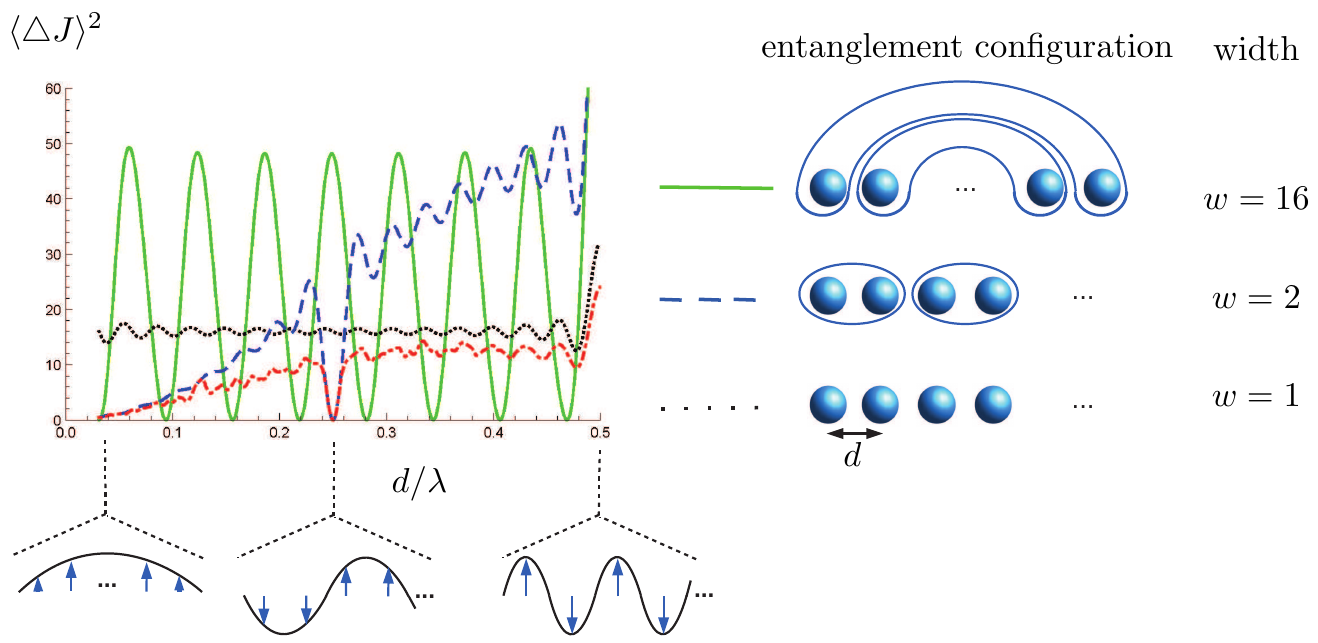}
   \end{center}
\caption{  Variance of the observable $\vec J$ for different parameters $\lambda$ and entanglement configurations for $N=16$ particles. Pairs of encircled particles form together the state $\ket{\psi^-}$. The product state (non-encircled particles, $w=1$) is chosen in such a way, that it minimizes the variance. Although, both entangled states possess the same entanglement depth $k=2$, they exhibit quite different behavior due to their different spatial distribution of entanglement. The red dashed-dotted line corresponds to the $w=2$ entanglement configuration (blue dashed line) optimized over all states with the given entanglement configuration.}
\label{fig:sin_B}
\end{figure}

The entanglement depth $k$ is an important variable characterizing a state as long as no spatial dependencies are involved. Though, the size of experimental available quantum systems has grown in recent times. As a consequence, the approximation that our quantum system under investigation couples only to spatially constant fields  becomes weaker. Furthermore, the spatial distribution of entanglement helps also to investigate quantum phase transitions \cite{Osterloh2002, Osborne2002,Gu2004,Biswas2014,Richerme2014b} or to distinguish different ground states of the generalized Heisenberg spin-chain \cite{Eckert2008,Chiara2011}. 

An example of a position depended field coupled to a quantum systems is given e.g. by a magnetic gradient field inducing a position dependent rotation of the spin as depicted in \fig{fig:spin_rotation}. The projection of the spin leads to the sinusoidal observable
\BE
\vec{B}=\sum\limits_j \sin\left(2\pi \frac{x_j}{\lambda}\right) \vec{\sigma}_j
\EE
with the position $x_j$ of particle $j$ and the wavelength $\lambda$ determined by the gradient and the interaction time. The variance $(\triangle \vec{B})^2=(\triangle B_x)^2+(\triangle B_y)^2+(\triangle B_z)^2$ can be used to determine the entanglement of the state similar to the variance of the total spin. However, $(\triangle \vec{B})^2$ not only depends on the amount of entanglement but also on its spatial distribution. Assume for example a state $\ket{\Psi}=\bigotimes_{\lbrace j,k\rbrace}\ket{\psi^-}_{j,k}$ consisting of pairs of particles in the state $\ket{\psi^-}_{j,k}=(\ket{01}_{j,k}-\ket{10}_{j,k})/\sqrt{2}$. The variance of this state now depends on the spatial distribution of the $\ket{\psi^-}_{j,k}$ states as shown in \fig{fig:sin_B}. Whereas $(\triangle J)^2$ is equal to zero for all configurations, $(\triangle B)^2$ is only equal to zero if the entangled pairs $\ket{\psi^-}_{j,k}$ are situated in such a way, that the coupling strength $a_j = \sin (2\pi x_j/\lambda)$ is equal for entangled particles $j$ and $k$. The variance becomes larger, the more the coupling strengths $a_j$ and $a_k$ differ from each other. This behavior is not only valid for the singlet state $\ket{\psi^-}$ but stays also valid if we minimize over all states. In general, we find the lower limit of the variance \cite{Woelk2015}
\BE
\underset{\ket{\psi}}{\text{min}}(\Delta \vec{B})^2_{(j,k)}=\left\lbrace \begin{array}{cl}
 a_j^2\left(2+2\varepsilon^2-\frac{4\varepsilon^2}{(1-\varepsilon)^2}\right)& -1\leq \varepsilon\leq \varepsilon_0\\
3a_j^2(1-\varepsilon)^2& \varepsilon_0\leq \varepsilon \leq 1
\end{array}
\right.
\EE
 where we assumed with out loss of generality  $|a_j|>|a_k|$ and defined $\varepsilon=a_k/a_j$ and $\varepsilon_0=2-\sqrt{3}\approx 0.27$. Here, we want to note that the state minimizing $(\Delta \vec{B})^2_{(j,k)}$ is given by the singlet state $\ket{\psi^-}$ for $\varepsilon>2-\sqrt{3}$. In this way, the minimal variance for states, where only nearest-neighbor entanglement between particles $j$ and $j+1$ for odd $j$ is allowed, can be determined. As can be seen in \fig{fig:sin_B} this bound (red dashed-dotted line) is valid for the product states and the nearest-neighbor configuration of $\ket{\psi^-}$ but is violated by other spatial configurations.

\begin{figure}[t]
\includegraphics[width=0.8\textwidth]{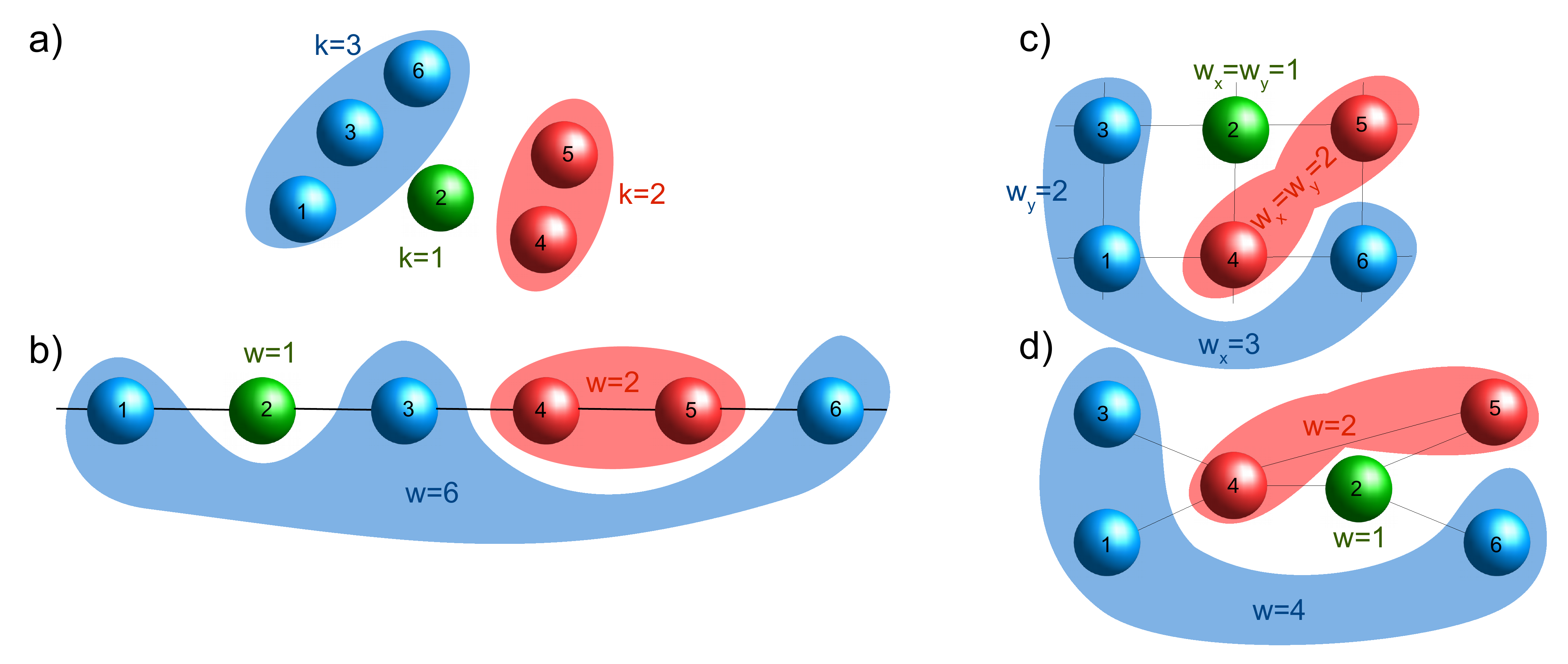}\\
 \caption{Comparison of entanglement depth (a) and entanglement 
 width in 1D (b) and 2D (c) or a graph (d) of the state $\ket{\Psi}=\ket{\psi_{1,3,6}}\otimes \ket{\psi_{4,5}}\otimes\ket{\psi_{2}}$: whereas the entanglement 
 depth disregards any spatial ordering, the definition of entanglement 
 width requires the particles to be spatially ordered, e.g. in a 
 spin chain or a grid. The entanglement depth of the state  $\ket{\Psi}$ in (a)
 is given by $k=3$ (since maximally three particles are entangled). This
 is a lower bound on the entanglement width in (b), which  equals $w=6$
 (since entanglement occurs over a distance of six particles in the chain). In the case of the 2d grid in (c) we distinguish between the width in the horizontal direction ($w_x=3$) and vertical direction ($w_y=2$). In a general graph (d) where edges denote possible interaction between two particles there might exist different ways to connect two particles. In this case, we take always the shortest way to determine the entanglement width \cite{Woelk2015}.}
 \label{fig:ent_width}
\end{figure}

Similar to multipartite entanglement, the exact investigation of the spatial distribution of entanglement is very resource intensive or even impossible for many-particle states. Hence,  we define instead the  width of entanglement (see \fig{fig:ent_width}) as a characteristic measure of the spatial distribution of entanglement similar to the entanglement depth for multipartite entanglement.

To define the width of entanglement $w$ we need to define a spatial ordering given e.g. by a linear chain of trapped ions (\fig{fig:ent_width} b)), a lattice of cold atoms (\fig{fig:ent_width} c)) or a graph with given interactions (\fig{fig:ent_width} d)). The distance between two particles is then determined by following the given structure and counting  particles in between. As a consequence, the width of entanglement $w$ of a pure state $\ket{\Psi}=\bigotimes_j \ket{\psi_j}$ is defined as the maximal distance $w$ of two entangled particles within the states $\ket{\psi_j}$.  A completely separable state exhibits an entanglement width of $w=1$. The entanglement width of a mixed state is defined by the minimum with $w$ over all decomposition $\varrho=\sum_k p_k \ket{\psi_k}\bra{\psi_k}$, that is 
 \BE
 w(\varrho)=\underset{\textrm{decompositions}}{\textrm{min}}\Big[\underset{k}{\textrm{max }}\{w(\psi_k)\}\Big].
 \EE
By definition, the entanglement depth is a lower bound of the entanglement width. However, the entanglement width does not make any statement about the entanglement depth.

To use the variance $(\triangle B)^2$ to detect the width of entanglement, we need to not only optimize over the state  but to also find the optimal pairing. For example, for the width $w=2$ ion $j$ might be entangled with $j-1$ or $j+1$ or not entangled at all. Still, these is a classical optimization problem which for some configurations is easy to solve or can be at least approximated \cite{Woelk2015}. For example, for $N$ particles situated at $x_j=x_0+j\cdot d$ with $d=\lambda/(2N)$ and $x_0=d/2$ the optimal pairing for only nearest-neighbor entanglement is given by entangling all odd particles $j$ with their right neighbor $j+1$. As a consequence, the variance $(\triangle B)^2$ for states with entanglement width $w\geq2$ are bounded from below by
\BE
(\Delta \vec{B})^2\geq\frac{3}{2} N [1-\cos(\frac{\pi}{N})]\approx \frac{3\pi^2}{4N}.\label{eq:rn}
\EE
In this way, the width of entanglement can be determined without addressing of single subsystems, solely with global measurements.

\section{Conclusion}

In summary, we have discussed in this lecture note different quantum properties such as the wave-particle duality and entanglement. However, both phenomena are based like many other quantum properties on non-commuting observables. As a result, measurement procedures to observe wave-particle duality or entanglement are based on similar principles such as the Cauchy-Schwarz inequality. The only difference is that to observe wave-particle duality or the Heisenberg uncertainty, we apply these principles on the whole system whereas we apply them to single subsystems to observe entanglement.

I thank M.S. Zubairy, O. G\"uhne and W.P. Schleich for fruitful discussions on these topics and D. Kiesler, S. Altenburg, T. Kraft, H. Siebeneich, P. Huber and J. Hoffmann for their careful reading of the manuscript.

\begin{appendix}
\section{Proof of \eq{eq:exp_schmidt}\label{app:exp_schmidt}}
A separable state can be decomposed into $\rho=\sum_n p_n \ket{\alpha_n}\bra{\alpha_n} \otimes\ket{\beta_n}\bra{\beta_n}$ with $\sum_n p_n=1$ and $\ket{\alpha_n}$,$\ket{\beta_n}$ being arbitrary normalized states of system $A$ and $B$, respectively. The expectation values of the joined observables $M_{j,k}$ factorize for the product states $\ket{\alpha_n}\ket{\beta_n}$. Therefore, the sum of the observables can be rewritten by
\BE
\sum\limits_{j,k=1}^3 \langle M_{j,k}\rangle_\sep = \sum\limits_n p_n \bra{\alpha_n}(\sigma_x+\sigma_y+\sigma_z)\ket{\alpha_n}\bra{\beta_n}(\sigma_x+\sigma_y+\sigma_z)\ket{\beta_n}
\EE
for separable states. Since $\langle \sigma_x+\sigma_y+\sigma_z\rangle \leq 1$ and $\sum_n p_n=1$ we find
\BE
\sum\limits_{j,k=1}^3 \langle M_{j,k}\rangle_\sep \leq 1.
\EE
\end{appendix}
 


\end{document}